\documentclass[useAMS,usenatbib]{mn2e}
\usepackage{amsfonts}
\usepackage{txfonts}
\usepackage{mathrsfs}
\usepackage{amssymb}
\usepackage{psfig}
\usepackage{graphicx,times}
\usepackage{natbib}

\title[massive star formation of the RMS]{Molecular line study of massive star forming regions from the RMS survey}
\author[Naiping Yu and Jun-Jie Wang]{Naiping Yu$^{1,2,3}$\thanks{E-mail: yunaiping09@mails.gucas.ac.cn} and
Jun-Jie Wang$^{1,2}$\footnotemark[1]\thanks{}\\
$^{1}$National Astronomical Observatories, Chinese Academy of
Sciences, Beijing 100012, China\\
$^{2}$NAOC-TU Joint Center for Astrophysics, Lhasa 850000, China\\
$^{3}$Graduate School of the Chinese Academy of Sciences, Beijing
100080, China }
\begin{document}

\date{Accepted 20** December 15. Received 20** December 14}

\pagerange{\pageref{firstpage}--\pageref{lastpage}} \pubyear{2002}

\maketitle

\label{firstpage}

\begin{abstract}
In this paper we selected a sample of massive star forming regions
from the Red MSX Source (RMS) survey, to study star formation
activities (mainly outflow and inflow signatures). We focused on
three molecular lines from the Millimeter Astronomy Legacy Team
Survey at 90 GHz (MALT90): HCO$^+$(1-0), H$^{13}$CO$^+$(1-0) and
SiO(2-1). According to previous observations, our sources could be
divided into two groups: nine massive young stellar object (MYSO)
candidates (radio-quiet) and ten HII regions (having spherical or
unresolved radio emissions). Outflow activities were found in eleven
sources while only three show inflow signatures in all. The high
outflow detection rate means outflows are common in massive star
forming regions. The inflow detection rate was relatively low. We
suggest this was due to beam dilution of the telescope. All the
three inflow candidates have outflow(s). The outward radiation and
thermal pressure from the central massive star(s) do not seem to
strong enough to halt accretion in G345.0034-00.2240. Our simple
model of G318.9480-00.1969 shows it has an infall velocity of about
1.8 km s$^{-1}$. The spectral energy distribution (SED) analysis
agrees our sources are massive and intermediate-massive star
formation regions.
\end{abstract}


\section{Introduction}
Massive stars have a deep impact on the evolution of galaxies.
Despite of their short lives, they determine the main chemical and
physical properties of the nearby interstellar medium (ISM). Both
theories and observations demonstrate they could hamper or trigger
the next generation of star formation, by the expansion of HII
regions and supernova explosions at the end of their lives (e.g.
Elmegreen \& Lada 1977; Lefloch \& Lazareff 1994). However, their
formation mechanism is still poorly understood as they used to form
in clusters, and detail study of massive star formation is further
hampered by their short lives, far distances, rare sources and
extensive dust extinctions. Several mechanisms have been proposed
that high-mass stars could form by accretion through massive disks
(e.g. Keto et al. 2002), competitive accretions in dense clusters
(Bonnell et al. 2004), ionized accretion (Keto $\&$ Wood, 2006), and
mergers of several low-mass stars (Bonnell et al. 1998). The former
three models are similar to that of low-mass star formation,
accompanied by outflow during the process of gravitational collapse.
Recent observations appear to favor the former three mechanism as
stellar mergers require a high stellar density of $\geq$ 10$^8$
stars pc$^{-3}$, which is more than 4 orders of magnitude larger
than those found in young embedded dense clusters (Bonnell 2002).
Given the accretion timescales in massive star-forming regions are
much shorter than those in low mass star-forming regions, the inside
nuclear burning of hydrogen takes place while massive stars are
still accreting, which means there is no pre-main-sequence stage for
massive stars. Then substantial UV photons and ionized stellar winds
rapidly ionize the surrounding hydrogen, forming a hyper-compact HII
region (HCHII) or ultracompact HII region (UCHII). Many questions
are still unclear: whether accretion could be halted by the strong
outward radiation and thermal pressure. Does it continue in an
ionized form? Does it continue through a molecular or ionized disk?
\begin{table*}
\begin{minipage}{13cm}
 \caption{\label{tab:test}List of our sources}
 \begin{tabular}{lclclclclclclcl}
  \hline
  \hline
 MSX  & RA  & Dec.  & Type & V$_{lsr}$$^a$ &D$^a$  & H$_2$O   \\

 name    & (J2000) & (J2000) &(candidate) &(km s$^{-1}$) & (kpc) & maser  &
 \\
  \hline
G316.5871-00.8086 & 14:46:23.24 &-60:35:47.0 & MYSO  & -45.8 &3.2 & Y \\
G318.9480-00.1969 & 15:00:55.31 &-58:58:52.6 & MYSO  & -34.5 &2.6 & Y \\
G326.4755+00.6947 & 15:43:18.94 &-54:07:35.4 & MYSO  & -41.6 &1.8 & Y \\
G326.5437+00.1684 & 15:45:53.26 &-54:30:01.3 & MYSO  & -74.1 &3.6 & N \\
G326.6618+00.5207 & 15:45:02.84 &-54:09:03.0 & MYSO  & -39.1 &1.8 & Y \\
G326.7796-00.2405 & 15:48:55.20 &-54:40:37.7 & MYSO  & -64.7 &3.6 & Y \\
G329.0663-00.3081 & 16:01:09.93 &-53:16:02.3 & MYSO  & -42.6 &2.9 & Y \\
G333.3151+00.1053 & 16:19:29.00 &-50:04:40.9 & MYSO  & -47.4 &4.0 & Y \\
G345.5043+00.3480 & 17:04:22.87 &-40:44:23.5 & MYSO  & -17.8 &2.1 & Y \\
\hline
\hline
G327.9018+00.1538 & 15:53:10.87 &-53:39:58.1 & HII  & -93.1 &5.8 & N \\
G329.3371+00.1469 & 16:00:33.13 &-52:44:47.1 & HII  & -107.1 &7.3 & Y \\
G332.8256-00.5498 & 16:20:10.46 &-50:53:28.6 & HII  & -4.0 &0.19 & N \\
G336.9842-00.1835 & 16:36:12.42 &-47:37:57.5 & HII  & -75.1 &10.8 & Y \\
G337.0047+00.3226 & 16:34:04.73 &-47:16:29.2 & HII  & -62.8 &10.8 & N \\
G339.1052+00.1490 & 16:42:59.58 &-45:49:43.6 & HII  & -78.2 &4.9 & N \\
G340.2480-00.3725 & 16:49:29.97 &-45:17:44.4 & HII  & -50.3 &3.9 & N \\
G345.0034-00.2240 & 17:05:11.19 &-41:29:06.3 & HII  & -28.7 &2.9 & N \\
G345.4881+00.3148 & 17:04:28.04 &-40:46:23.3 & HII  & -17.6 &2.1 & N \\
G348.8922-00.1787 & 17:17:00.10 &-38:19:26.4 & HII  & 1.0 &18.2 & Y \\

  \hline
 \end{tabular}
 \label{tb:rotn}

 a: Urquhart et al. (2007b; 2008)
\end{minipage}
\end{table*}

In recent years, using color selection criteria and the IRAS point
source catalogue, several attempts have been taken to search for
MYSOs (e.g. Molinari et al. 1996; Sridharan et al. 2002). However,
because of the large IRAS beam ($\sim$3-5$^\prime$ at 100 $\mu$m ),
these selected samples tend to be biased towards bright, isolated
sources and avoid dense clustered environments at the Galactic
mid-plane. By comparing the the colors of sources from the MSX and
2MASS point sources to those known MYSOs, Lumsden et al. (2002)
identified approximately 2000 MYSO candidates. The Red MSX Source
(RMS) survey is an ongoing multi-wavelength observational programme
and will provide us the largest MYSO sample for statistical studies.
Using the Australia Telescope Compact Array (ATCA), Urquhart et al.
(2007a) completed the 5 GHz observations of 892 RMS sources in the
southern sky. This programme divided these sources into three
groups: real MYSO candidates, HII regions (UCHII and HCHII) and
others such as evolved stars and planetary nebulae (PNe). To obtain
kinematic distances, Urquhart et al. (2007b; 2008) made $^{13}$CO
(1-0) and (2-1) observations at Mopra, Onsala and Purple Mountain
Observatory (PMO) 13.7 m telescope, and the 15 m James Clerk Maxwell
Telescope (JCMT), as well as archival data extracted from the
Galactic Ring Survey (GRS). They found that the majority of these
sources have multiple velocity components along each line of sight.
The multiple emission features make it difficult to assign a unique
kinematic velocity to each source. In order to identify a more
reliable molecular component, they further searched archival water
and methanol masers catalogues of Valdettaro et al. (2001) and
Pestalozzi et al. (2005), less abundant but denser gas molecular
traces like CS (2-1) observations by Bronfman et al. (1996). And
then  by using the rotation curve of Brand and Blitz (1993) and
their radial velocities, kinematic distances for all detected
components can be derived. Based on the above observations and
research, we selected about twenty RMS sources to study star
formation activities (mainly outflow and inflow signatures) using
data from Millimeter Astronomy Legacy Team 90 GHz survey (MALT90).
We present the introductions of our data and source selections in
section 2, results and analysis in section 3, and summary in section
4.

\section{Data and source selections}
The MALT90 is a large international project aimed at characterizing
the sites within our Galaxy where high-mass stars will form.
Exploiting the unique broad frequency range and fast-mapping
capabilities of the Mopra 22-m telescope, MALT90 maps 16 emission
lines simultaneously at frequencies near 90 GHz. These molecular
lines will probe the physical, chemical, and evolutionary states of
dense high-mass star-forming cores.We focused on three molecular
lines from the MALT90 Survey: HCO$^+$(1-0), H$^{13}$CO$^+$(1-0) and
SiO(2-1). HCO$^+$ often shows infall signatures and outflow
wings(e.g., Rawlings et al. 2004; Fuller et al. 2005).
H$^{13}$CO$^+$(1-0) provides optical depth and line profile
information. SiO (2-1) is seen when SiO is formed from shocked dust
grains, typically in outflows (Schilke et al. 1997). The survey
covers a Galactic longitude range  of $\sim$ -60 to $\sim$
15$^{\circ}$ and Galactic latitude range of -1 to +1 $^{\circ}$. The
observations were carried out with the newly upgraded Mopra
Spectrometer (MOPS). The full 8 GHz bandwidth of MOPS was split into
16 zoom bands of 138 MHz each providing a velocity resolution of
$\sim$ 0.11 km s$^{-1}$. The angular resolution of Mopra is about 38
arcsec, with beam efficiency between 0.49 at 86 GHz and 0.42 at 115
GHz (Ladd et al. 2005). The maps were made with 9$^{\prime\prime}$
spacing between adjacent rows. More information about this survey
can be found in Foster et al. (2011) and Hoq et al. (2013). The
MALT90 data includes (\emph{l}, \emph{b}, \emph{v}) data cubes and
(\emph{l}, \emph{b}) moment and error maps, and is publicly
available from the MALT90 Home Page\footnote{See
  http://atoa.atnf.csiro.au/MALT90}. The data
processing was conducted using CLASS (Continuum and Line Analysis
Single-Disk Software) and GreG (Grenoble Graphic) software packages.

In order to study massive star formation activities (mainly outflow
and inflow signatures), we selected nineteen sources from the RMS
survey by applying the following criteria: (1) Sources should not be
on the edge of known HII or supernova regions, considering the large
beam of the 22 m Mopra telescope; (2) According to the observations
of ATCA (Urquhart et al. 2007a), sources should be radio quiet or
have a simple spherical/unresolved radio emissions; (3) All sources
should be detected by MALT90; (4)The LSR velocities of HCO$^+$ and
H$^{13}$CO$^+$ should be the same as Urquhart et al. (2007b; 2008).
We need to mention here that our sample is not complete according to
the above four criteria. We wish a much more complete research of
the RMS sources by the MALT90 data in the near future. The
information of our selected sources are listed in table 1. Nearly
all of the MYSO candidates are associated with water masers
according to former observations, however in the HII regions only
three sources are associated with water maser emissions. This
suggests in massive star formation regions, water masers are more
likely to be associated with MYSOs than HII regions.
\begin{table*}
\begin{minipage}{13cm}
 \caption{\label{tab:test}Outflow parameters}
 \begin{tabular}{lclclclclcl}
  \hline
  \hline
 Source & Shift & $\Delta$$\upsilonup$ (Km s$^{-1}$)& N (HCO$^+$) ($\times$ 10$^{12}$ cm$^{-2}$) & M
 (M$_{\odot}$) & P$_{out}$ ($M_\odot$ km s$^{-1}$) & E$_{out}$ (M$_\odot$ [km s$^{-1}$]$^2$)
\\
  \hline
G326.4755+00.6947 & red &(-40, -33)& 22.4 & 16.8&118&412\\
                  & blue &(-47, -42)& 17.7& 19.0&95&238\\
  \hline
G326.7796-00.2405 & red &(-64, -62)&10.0 & 21.4&43&43\\
                  & blue &(-69, -66)& 8.91 & 23.9&72&108\\
  \hline
G333.3151+00.1053 & red &(-45.3, -44)& 4.8 & 15.2&20&13\\
                  & blue &(-50, -47)& 11.8 & 46.8&140&211\\
  \hline
G345.5043+00.3480 & red &(-12, -15)&12.2 & 6.7&20&30\\
                  & blue &(-20, -24)& 9.1& 10.0&40&80\\
  \hline
  \hline
G329.3371+00.1469 & red &(-106, -104)&8.9 & 78.4&157&157\\
                  & blue &(-112, -110 )& 5.6& 49.3&99&99\\
  \hline
G332.8256-00.5498 & red &(-49 -53)&14.0 & 1.0&4&8\\
                  & blue &(-65, -61)& 11.3& 1.7&7&14\\
  \hline
G340.2480-00.3725 & red &(-49, -46)&9.5 & 50.2&151&226\\
                  & blue &(-56, -53)& 10.7& 65.9&198&297\\
  \hline
G345.0034-00.2240$^a$ & red &(-22, -19)&6.1& 14.8&89&266\\
                  & blue &(-35, -31 )& 9.8& 28.6&172&515\\
  \hline
G345.4881+00.3148 & red &(-15, -12)&14.0 & 21.6&65&97\\
                  & blue &(-22, -19)& 21.8& 28.6&86&129\\
  \hline

 \end{tabular}
  \label{tb:rotn}

 a: The outflow parameters of this source are calculated by
 HNC(1-0).
\end{minipage}
\end{table*}

\begin{table*}
\begin{center}
 \caption{\label{tab:test}SED parameters}
 \begin{tabular}{lclclclclcl}
  \hline
  \hline
 RMS  &Stellar mass  & Disk mass   & Envelope mass & Total luminosity& Best-fit\\

 sources    & [M$_{\odot}$] & [M$_{\odot}$] & [M$_{\odot}$] &
 [L$_\odot$] & model
 \\
  \hline
G318.9480-00.1969& 9.56 & 4.68 $\times$ 10$^{-3}$ & 9.65 $\times$ 10$^{0}$& 5.72 $\times$ 10$^3$ &3014991\\
G326.4755+00.6947& 10.10& 9.77 $\times$ 10$^{-3}$ & 1.14 $\times$ 10$^{3}$& 9.06 $\times$ 10$^3$ &3000136\\
G326.7796-00.2405& 8.08 & 4.58 $\times$ 10$^{-2}$ & 2.68 $\times$ 10$^{1}$& 7.33 $\times$ 10$^2$ &3004818\\
G329.0663-00.3081& 9.49 & 5.52 $\times$ 10$^{-2}$ & 3.08 $\times$ 10$^{1}$& 1.45 $\times$ 10$^3$ &3010080\\
G333.3151+00.1053& 13.91& 5.88 $\times$ 10$^{-4}$ & 6.46 $\times$ 10$^{2}$& 1.70 $\times$ 10$^4$ &3000244\\
G345.0034-00.2240& 9.67 & 7.58 $\times$ 10$^{-2}$ & 4.42 $\times$ 10$^{1}$& 7.81 $\times$ 10$^3$ &3003254\\
  \hline
 \end{tabular}

 \end{center}
\end{table*}

\section{Results and Discussions}
Infall and outflow are two of the most important elements to
understand the theories of massive star formation. Infall can act to
replenish disk material as mass falls onto a protostar, while
outflows serve as a release mechanism for the angular momentum that
builds up during the accretion process. These motions can be studied
by investigating the profiles of optically thick and optically thin
molecular lines. Blue profile, a combination of a double peak with a
brighter blue peak or a skewed single blue peak in optically thick
lines, can be used to study infall motions (Sun et al. 2008). Surely
blue profile may also be caused by rotation and outflow. However,
infall motion is the only process that would produce consistently
the blue profile. Outflow and rotation only produce a blue
asymmetric line profile along a particular line of sight to a source
(Sun et al. 2008).

SiO is also a well-known tracer of recent outflow. In the cold
diffuse ISM, the element of Si is regarded to be frozen into dust
grains. When the gas is shocked (i.e., the gas through which a
protostellar outflow is passing) the dust grains can sublimate and
Si is released into gas phase. Thus, the detections of SiO emissions
from MYSOs could always be equal to the detections of recent outflow
activities. In the following sections, we use position-velocity (PV)
diagram of HCO$^+$ (1-0) and/or SiO (2-1) (if detected) to study
outflows in our sample. For G318.9480-00.1969, we also employed the
three dimensional radiative-transfer code RADMC3D, developed by C.
Dullemond\footnote{See
  http://www.ita.uni-heidelberg.de/dullemond/software/radmc-3d/},
to compute the HCO$^+$ (1-0) line emission from an infalling model.

\subsection{Outflow signatures detected by HCO$^+$ (1-0) and SiO (2-1)}
Outflow makes an important contribution to the line wing emissions
of HCO$^+$ (1-0), as it becomes optically thick quickly and readily
self-absorbs in dense gas regions. We drew PV diagrams for our
sources. According to the PV diagrams, we selected the integrated
range of wings and determined the outflow intensities of red and
blue lobes. Four HCO$^+$ (1-0) PV diagrams of our nine MYSO
candidates and four in ten HII regions show distinct wing emissions
(figure 1 and figure 2, respectively), which may be probably caused
by bipolar outflows. Figure 3 and figure 4 show the integrated maps
by integrating wing emissions of our MYSOs and HII regions,
respectively. The Mopra beam size is 38$^{\prime\prime}$ and the
spacing of the spectra is 9$^{\prime\prime}$, so there are about 4
$\times$ 4 = 16 spectra within a single beam. In the sample of
MYSOs, the peak emissions of red and blue lobes are within one beam
and we could not determine the outflow directions. It also seems
impossible that all the axis of these outflows are nearly parallel
with the line of sight. By comparing figure 3 and figure 4, it can
be noted that the MYSO outflows are more compact than those found in
HII regions. Considering the mean distance of our RMS sources with
radio emission ($\sim$ 5.7 kpc) is larger than those radio quiet
($\sim$ 2.6 kpc), our result agrees MYSO candidates should be in
much earlier stage than HCHII/UCHII regions.

Figure 5 shows the five detected SiO spectra toward our sources.
Among these five sources, three (G318.9480-00.1969,
G326.4755+00.6947, G329.0663-00.3081) are MYSO candidates and the
other two (G340.2480-00.3725 and G345.0034-00.2240) are HII regions.
Models and observations also suggest SiO emissions could be caused
by photodissociated region (PDR) and not an outflow (e.g. Schilke et
al. 2001; Shepherd et al. 2004). For the MYSO candidates, as they
are radio quiet, SiO emissions triggered by PDR around HII regions
are impossible. However, a PDR could also exist without the presence
of ionized gas. Through the method described by Klaassen et al.
(2007), we estimated the SiO column densities $>$ 10$^{14}$
cm$^{-2}$ of the five sources, much larger than those found in PDRs
($\sim$ 10$^{12}$ cm$^{-2}$) (Schilke et al. 2001). We thus regard
these SiO emissions are due to recent outflow activities.

We should mention that the outflow detection rate is just a low
limit. One reason is due to the beam dilution, especially for the
MYSO candidates. For example outflow activity was not found in MYSO
G318.9480-00.1969 through PV diagram analysis, however the SiO
emission discussed above shows recent outflow(s) in this region.
Outflow(s) in this source may be very young and is thus heavily
diluted. The other reason is due to the molecular line we used to
trace outflow. Considering the wide range of physical conditions in
star formation regions (i.e. gas densities from $\sim$ 10$^{-20}$ g
cm$^{-3}$ to $\sim$ 10$^{-16}$ g cm$^{-3}$, and gas temperatures
from $\sim$ 10 K to $\sim$ 100 K), there is non-trivial molecular
tracer to detect outflows in all conditions. For example, the PV
diagram of HCO$^+$ in G345.0034-00.2240 does not show red-shifted
emission wings(figure 6). The gas near -17 km/s is possibly
unrelated to this source. However, when we chose HNC instead of
HCO$^+$, wing emissions caused by bipolar outflow was evident.
Besides, the SiO spectra of G345.0034-00.2240 extends from -8 km
s$^{-1}$ to -48 km s$^{-1}$. The wide line emissions further suggest
young outflow activities. PV diagram of SiO in G345.0034-00.2240 cut
along east-west direction is shown in figure 7. Figure 7 also shows
the maps of the integrated blue and red shifted SiO (2-1) emission
(the dash contours). It can be noted SiO traces the inner region of
the outflow, compared to HNC. The outflow detection rate by SiO in
MYSOs ($\sim$ 33 $\%$) is much higher than that in HII regions
($\sim$ 20 $\%$). That is because SiO is particularly well suited to
tracing recent outflows, as it persists in the gas phase for only
$\sim$ 10$^4$ years after being released by shocks(e.g., Pineau des
Forets et al. 1997).

Assuming that HCO$^+$ (1-0) emission in the line wings to be
optically thin and Local Thermodynamic Equilibrium (LTE), X(HCO$^+$)
= [HCO$^+$]/[H$_2$] = 10$^{-8}$ (Turner et al. 1997) and T$_{ex}$ =
15 K, we derive the column density using:
\begin{equation}
N(HCO^+) = Q(T_{ex}) \frac{8 \pi v_0^3}{c^3} \frac{g_l}{g_u}
\frac{1}{A_{ul}} [1 - e^{-hv_0/kT_{ex}}]^{-1} \int \tau dv
\end{equation}
where $\nu$$_0$, \itshape g$_u$, g$_l$\upshape and A$_{ul}$ are the
rest frequency, the upper and lower level degeneracies and the
Einstein's coefficient of HCO$^+$, Q(T$_{ex}$) is the partition
function, and \itshape c \upshape is the speed of light. In
addition, by assuming that the HCO$^+$ emission is optically thin in
the line wings, we use the approximation:
\begin{equation}
\int \tau dv = \frac{1}{J(T_{ex}) - J(T_{bg})} \int T_{mb} dv
\end{equation}
where T$_{bg}$ is the temperature of the background radiation (2.73
K). Using $M_{out}$ = $\mu$ \itshape m$_H$ d$^2$ \upshape $\Omega$
X(HCO$^+$)$^{-1}$ $N$(HCO$^+$), we obtain the masses for the red and
blue molecular outflows, where $N$(HCO$^+$) is the HCO$^+$ column
density calculated through the above equations, \itshape d \upshape
is the distance, $\Omega$ is the area of the lobes (within 50$\%$ of
each peak emission), and \itshape m$_{H}$ \upshape is the hydrogen
atom mass. We adopt a mean molecular weight per H$_2$ molecule of
$\mu$ = 2.72 to include helium. We estimate the momentum and energy
of the red and blue lobes using
\begin{equation}
P_{out} = M_{out} V
\end{equation}
and
\begin{equation}
E_{out} =\frac{1}{2} M_{out} V^2
\end{equation}
where $V$ is a characteristic velocity estimated as the difference
between the maximum velocity of HCO$^+$ emission in the red and blue
wings respectively, and the molecular ambient velocity ($V_{lsr}$).
The derived parameters are shown in table 2.

\subsection{Infall }
\subsubsection{The blue profile}
Previous studies show HCO$^{+}$ (1-0) is a good inward motion tracer
in star formation regions (e.g. Sun et al. 2008). The five HCO$^+$
(1-0) emission lines of our sources (figure 8, right panels) are far
of having a simple Gaussian shape, presenting asymmetries, and
spectral wings or shoulders, which suggest that the molecular gas is
affected by the dynamics of these star-forming regions. However, we
know double peak could also be caused by two velocity components in
the line of sight. The detections of optically thin molecular lines
such as H$^{13}$CO$^+$ (1-0) could help us to rule out this
possibility. Mapping observations could help us to identify whether
this was caused by inward motions or other dynamics such as outflow,
rotation and expansions of HII regions. Figure 8 (the left panels)
shows the mapping observations towards our sources with double
peaked HCO$^+$ (1-0) emissions. All the spectra of H$^{13}$CO$^+$ in
figure 8 peaks near the dip of HCO$^+$. Besides, three of them
(G318.9480-00.1969, G345.0034-00.2240 and G345.4881+00.3148) show
consistent blue profiles, indicating infall motions. To quantify the
blue profile, we further used the asymmetry parameter $\delta$V
defined as the difference between the peak velocities of an
optically thick line V(thick) and an optically thin line V(thin) in
units of the optically thin line FWHM (Full Width at Half Maximum)
dV(thin): $\delta$V = $\frac{V(thick) - V(thin)}{dV(thin)}$.
Mardones et al. (1997) adopted a criterion $\delta$V $<$ -0.25 to
indicate blue asymmetry. Our calculations further demonstrate blue
profiles caused by inflow in G318.9480-00.1969($\delta$V = -0.27),
G345.0034-00.2240 ($\delta$V = -0.63) and G345.4881+00.3148
($\delta$V = -0.37). Outflow activities are also detected in these
regions (see section 3.1). This suggests that like low star
formation theories, massive stars in these regions are probably
forming through accretion-outflow process. For G345.0034-00.2240, if
the detection of recent outflow activity traced by SiO and the
appearance to be undergoing infall are indeed caused by the central
massive young star(s), this suggests the outward radiation and
thermal pressure from the central massive star(s) do not strong
enough to halt accretion. Like the case of G10.6-0.4 (Keto $\&$ Wood
2006), accretion flow in this region may be ionized. Given the low
angular resolution of the data (at 2.9 kpc, 38$^{\prime\prime}$ is
over half a parsec), it is also possible that a lower mass star is
forming in the vicinity. Deeper observations should be carried out
to study our speculation.

Sun et al. (2008) made single-pointing and mapping observations of
HCO$^{+}$ (1-0) from the 13.7m telescope of PMO. Among their 29
massive star-forming cores (mainly UCHII regions), six sources were
identified to be strong infall candidates. The detection rate is
consistent with our study, even though the resolution of Mopra is
much higher than that of the 13.7m telescope (38$^{\prime\prime}$
vs. 58$^{\prime\prime}$). This may be because all of their source
distances are less than 4 kpc while half of our sources locate more
than 5 kpc away. Klaassen et al. (2007,2012) also obtained JCMT
observations of HCO$^+$/H$^{13}$CO$^+$ (4-3) to trace large scale
inward motions in a sample of massive star-forming regions (mainly
MYSOs, HCHII and UCHII regions). The infall rate in our sample is
relatively low compared with their work. This partly may be due to
their higher resolution (15$^{\prime\prime}$ vs.
38$^{\prime\prime}$). Besides, the J = 4-3 transition is likely a
better asymmetry tracer than J = 1-0 (see figure 8 of Tsamis et al.
2008). Only one source in our MYSO candidates shows infall
signature. This may be because the infall area within MYSO is
relatively smaller than that in UCHII regions, and then more likely
to be beam diluted.

\subsubsection{A simple infall model of G318.9480-00.1969}
G318.9480-00.1969 is the only MYSO candidate showing infall
signature. In this section, in order to constrain the spatial and
dynamic structures of G318.9480-00.1969, we constructed a
radiative-transfer model that reproduces the observations. The
three-dimensional radiative-transfer code RADMC-3D, developed by C.
Dullemond, was employed to compute the dust temperature from stellar
heating and the continuum and line emission of an infall model. The
dust opacity is from Ossenkopf $\&$ Henning (1994) without grain
mantles or coagulation. The molecular data of HCO$^+$ comes from the
Leiden LAMDA database\footnote{
http://www.strw.leidennuniv.nl/$\sim$moldata}. The line transfer
assumes the gas temperature to be equal to the dust temperature and
LTE (full non-LTE radiation transfer is also planned for RADMC-3D).

In our model, the volume density follows a radial power law, $n$
$\propto$ $r^{-1.5}$, with a total mass of 31 M$_\odot$ (estimated
from our observations) within a box of (4000 AU)$^3$. Inside there
is a star of 10 M$_\odot$ (see section 3.3). The gas has a
turbulence velocity of 0.5 km s$^{-1}$ and is radially infalling
with 1.8 km s$^{-1}$ to the central star. The model is inevitably
simplified. A comparison of the observed and model spectra is shown
in figure 9. At the center, the model spectra matches well with the
observation. However, the observed line seems to have line wing
emissions, probably caused by outflow activities. The detected SiO
spectra in this region indeed implies beam diluted outflow(s). Even
though our model is consistent with the data, we should realize it
does not provide errors of the parameters, and different models may
also fit as well. However, it provides us a way to study the
structure of star forming regions.

\subsection{Spectral energy distribution}
In this section, we try to fit the spectral energy distribution
(SED) of our sources using the tool developed by Robitaille et al.
(2007). Briefly, the SED-fitting tool works as a regression method
to find the SEDs within a specified $\chi$$^{2}$ from a large grid
of models after fitting the input data points. The grid of models
contains stellar masses, disk masses, mass accretion rates, and
line-of-sight (LOS) inclinations. The grid of YSO models was
computed by Robitaille et al. (2006) using the 20,000
two-dimensional radiation transfer models from Whitney et al.
(2003a, 2003b, 2004). Each YSO model has SEDs for 10 viewing angles
(inclinations), so the total YSO grid consists of 200,000 SEDs. We
use archival data from 2MASS, IRAC of Spitzer, MSX to fit the SED of
our sources. In addition to the best-fit model (the black line in
figure 10), we also show the range of possible parameters that can
be derived from models within the range of $\chi$$^{2}$/$\nu$
-$\chi$$^{2}$$_{best}$/$\nu$ $\le$ 4 ($\nu$ represents the number of
data points). The best derived model parameters are listed in table
3 and the resulting SEDs are shown in figue 10. Of these six
sources, five are HII regions. This may be because the protostar(s)
of MYSOs are still deeply imbedded, even infrared emissions are hard
to be detected. The SED agrees our sources are massive and
intermediate-massive star formation regions, with masses ranging
from 8 to 14 M$_{\odot}$.

\section{Summary}
By analyzing HCO$^+$(1-0), H$^{13}$CO$^+$(1-0) and SiO(2-1)
molecular data from MALT90, we studied the outflow and infall
activities in nineteen RMS sources. The high outflow detection rate
($\ge$ 58 $\%$) suggests that outflows are common in massive star
forming regions as in low mass star forming regions. All of the
detected outflows in our radio-quiet RMS sources are much more
compact than those found in radio-loud RMS sources, indicating they
are at earlier stages. The outward radiation and thermal pressure
from the central massive star(s) of G345.0034-00.2240 do not seem to
strong enough to halt accretion in this region. The detection of
recent outflow activity traced by SiO and the appearance to be
undergoing infall in this region suggest ionized accretion flow can
continue through an HII region and massive star(s) in this region
could be formed through ionized accretions. Only one source of our
MYSO candidates shows infall signature. This may be because the
infall area with MYSO is relatively small, and thus more likely to
be beam diluted. A simple model of G318.9480-00.1969 shows it has an
infall velocity of about 1.8 km s$^{-1}$. The spectral energy
distribution (SED) further agrees our RMS sources are massive and
intermediate-massive star formation regions.

\section{Acknowledgements}
We thank the anonymous referee for constructive suggestions. This
paper made use of information from the Red MSX Source (RMS) survey
database http://rms.leeds.ac.uk/cgi-bin/public/RMS$_{-}$DATABASE.cgi
which was constructed with support from the Science and Technology
Facilities Council of the UK. This research made use of data
products from the Millimetre Astronomy Legacy Team 90 GHz (MALT90)
survey. The Mopra telescope is part of the Australia Telescope and
is funded by the Commonwealth of Australia for operation as National
Facility managed by CSIRO.

\begin{figure*}
\onecolumn \centering
\includegraphics[width=3in,height=3in,angle=270]{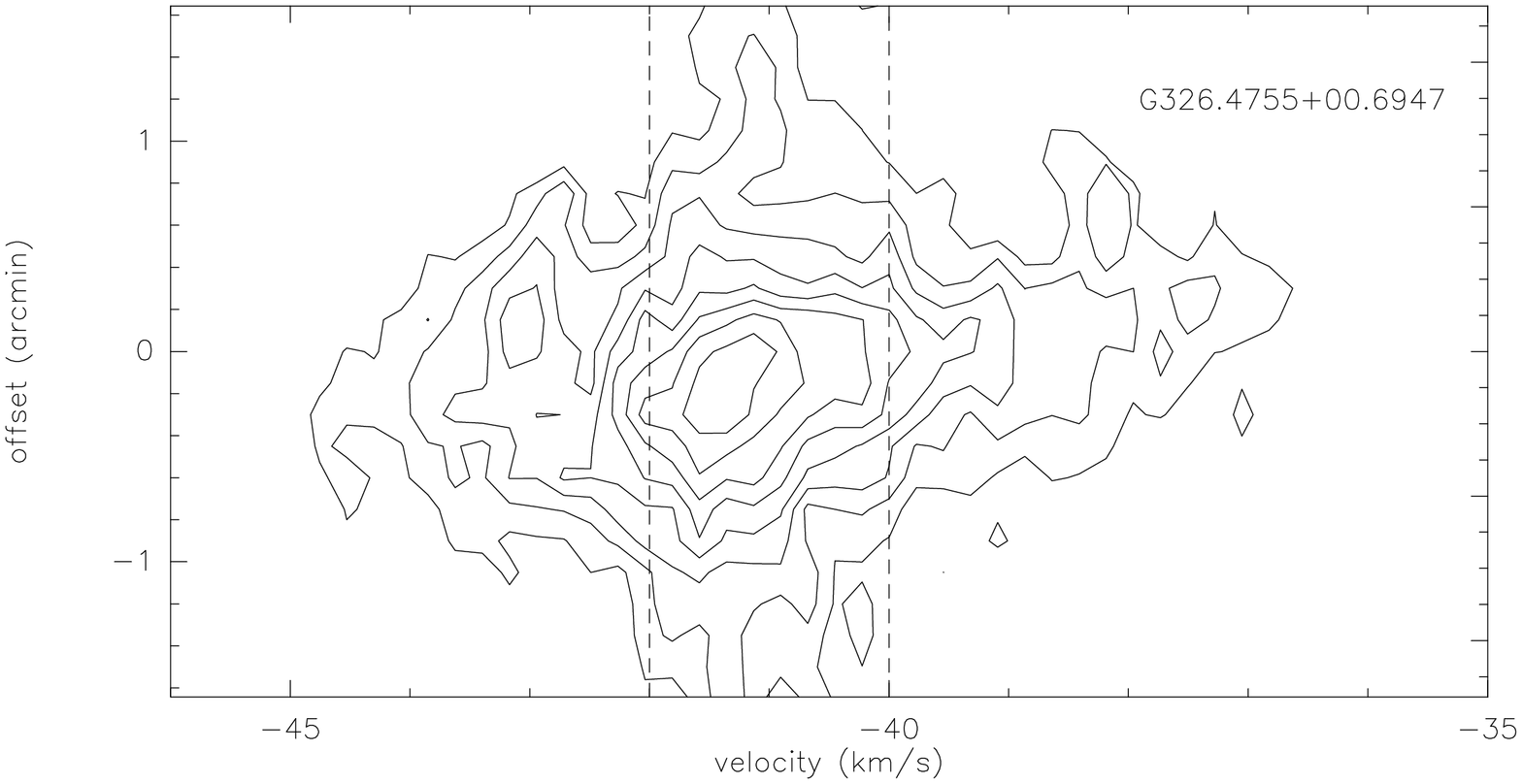}
\includegraphics[width=3in,height=3in,angle=270]{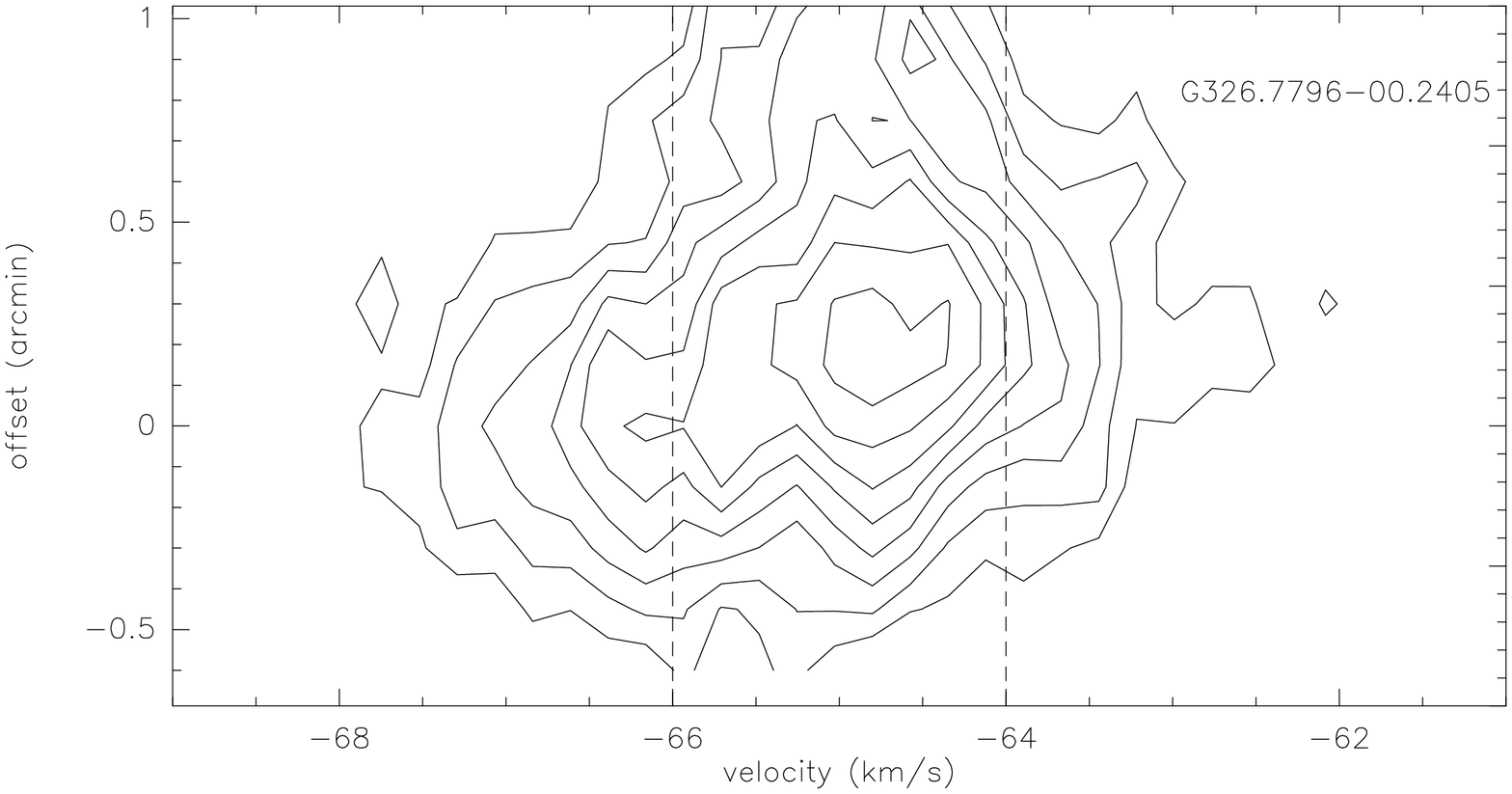}
\includegraphics[width=3in,height=3in,angle=270]{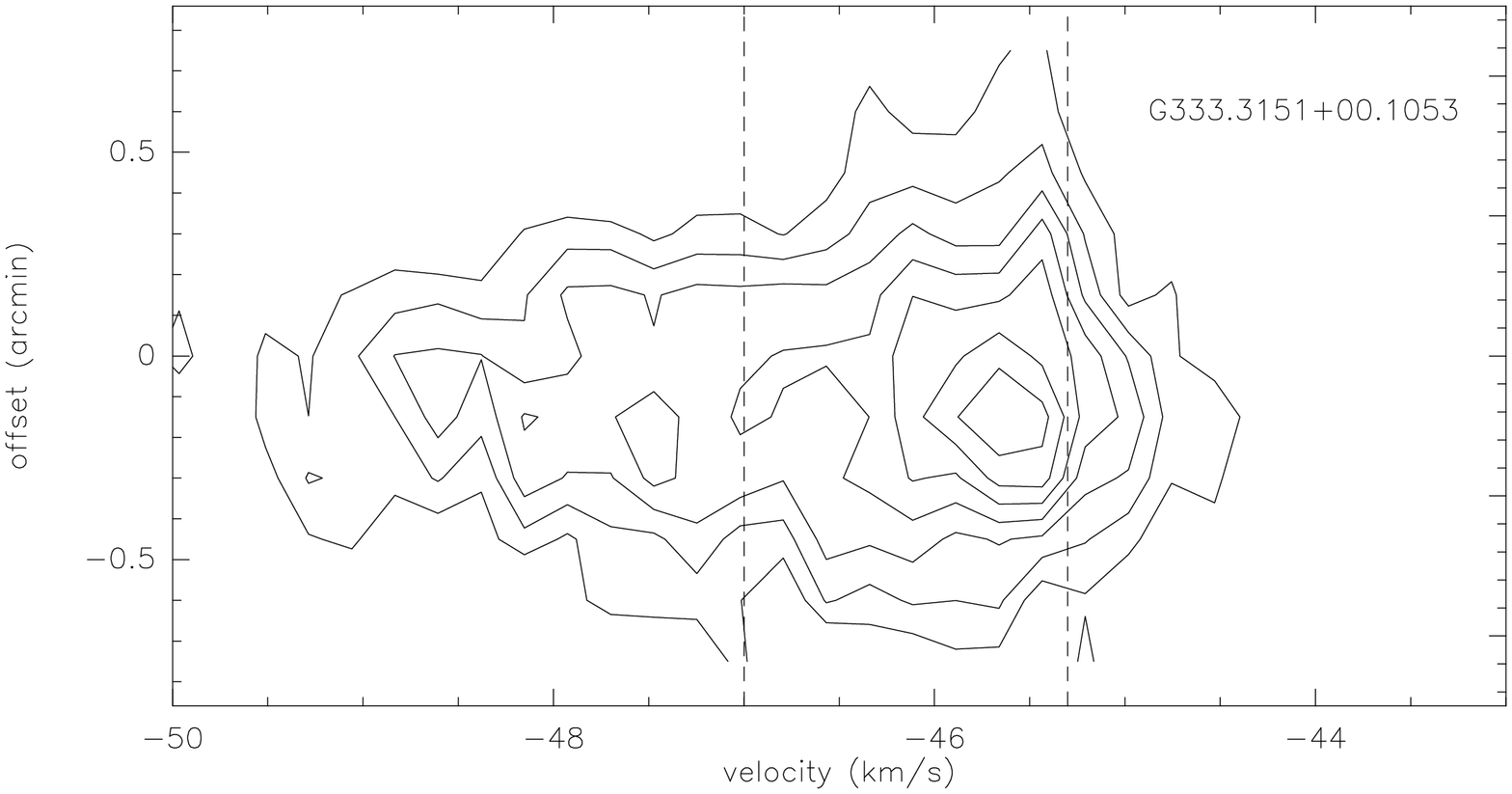}
\includegraphics[width=3in,height=3in,angle=270]{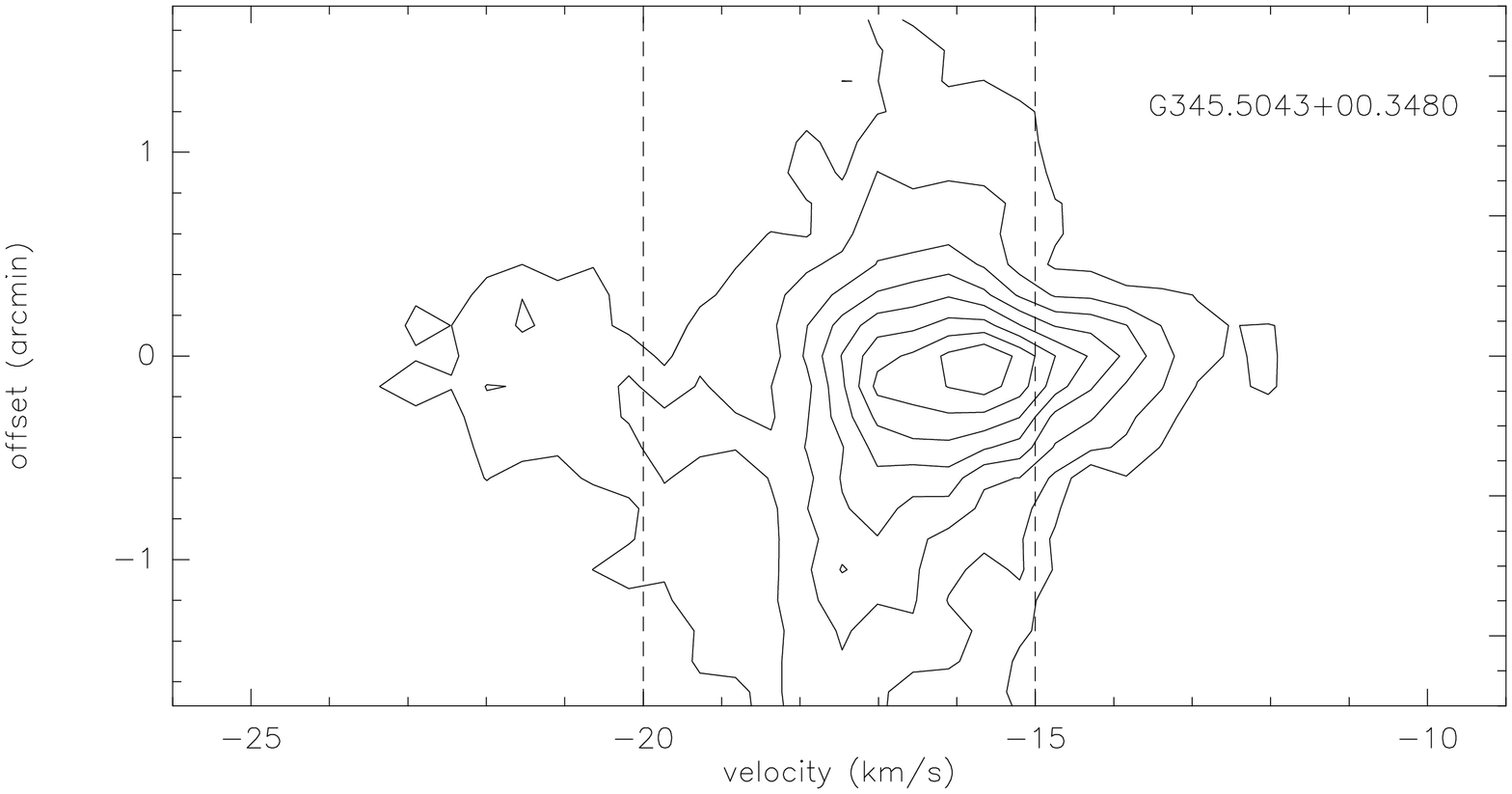}
\caption{ PV diagrams of the MYSO candidates show HCO$^+$ (1-0) wing
emissions. Contours are 20$\%$, 30$\%$...90$\%$ of the peak
emissions. The dashed lines indicate the velocity ranges for the
blue and red wings as listed in table 2.}
\end{figure*}

\begin{figure*}
\onecolumn \centering
\includegraphics[width=3in,height=3in,angle=270]{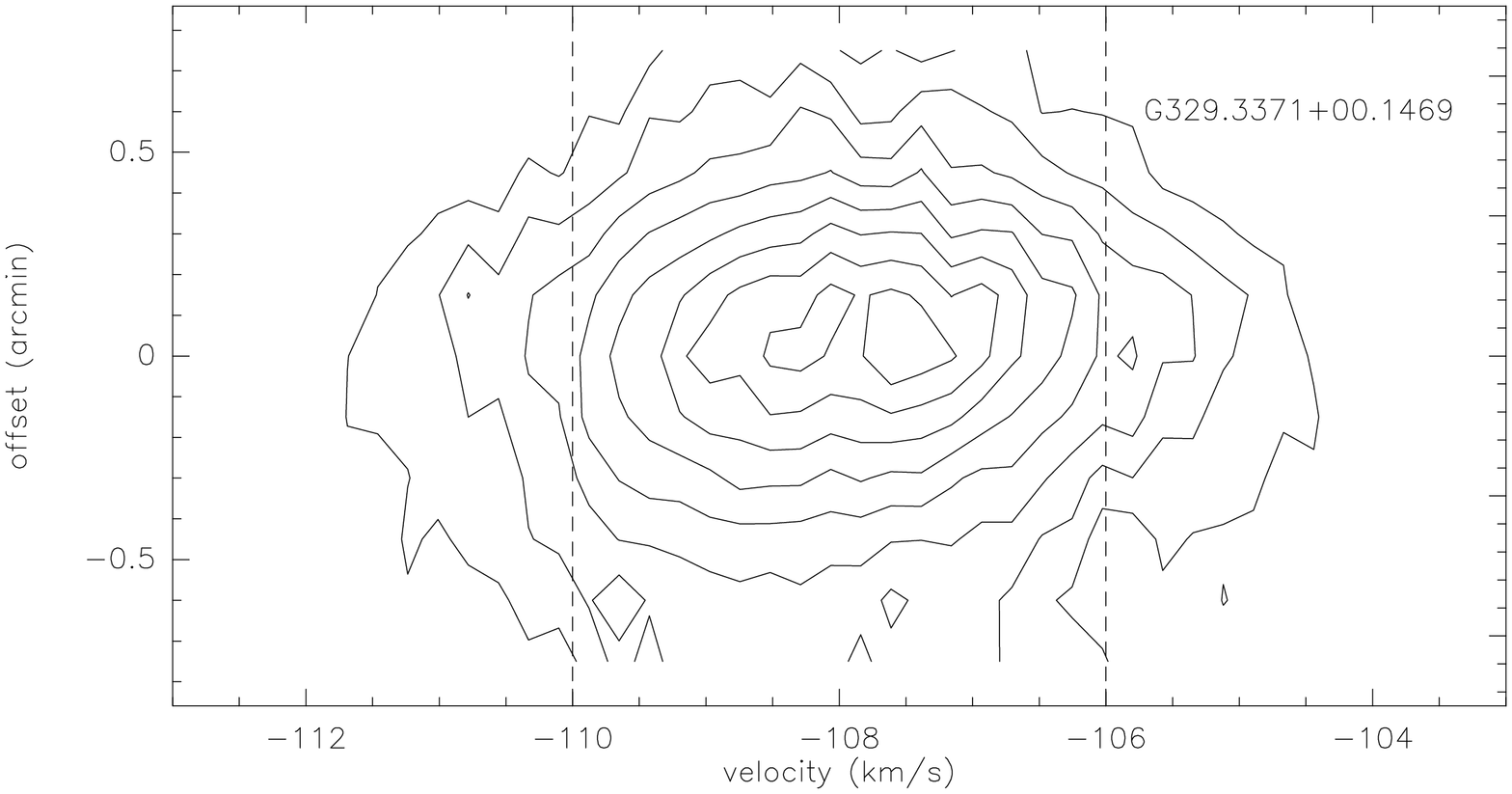}
\includegraphics[width=3in,height=3in,angle=270]{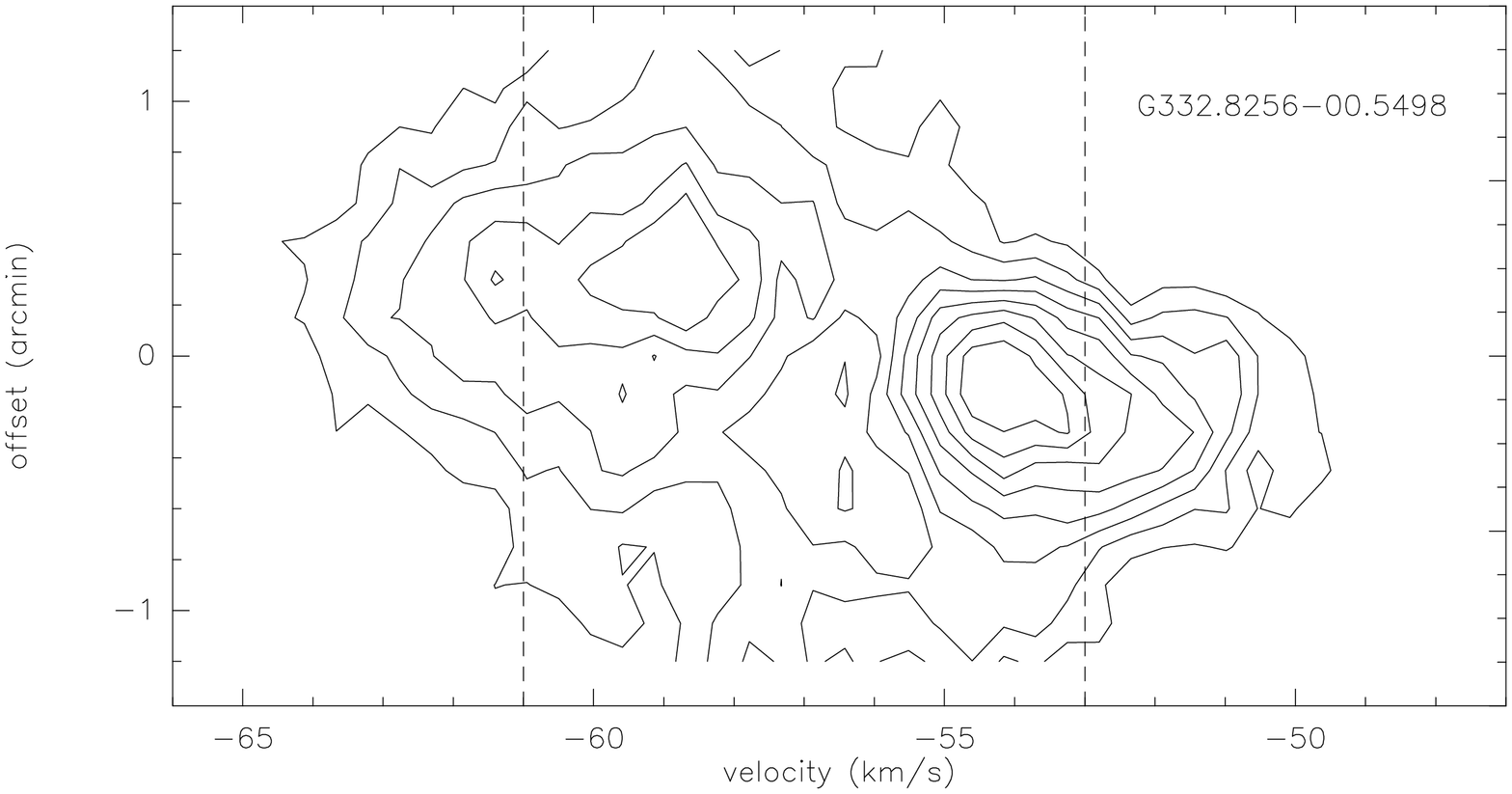}
\includegraphics[width=3in,height=3in,angle=270]{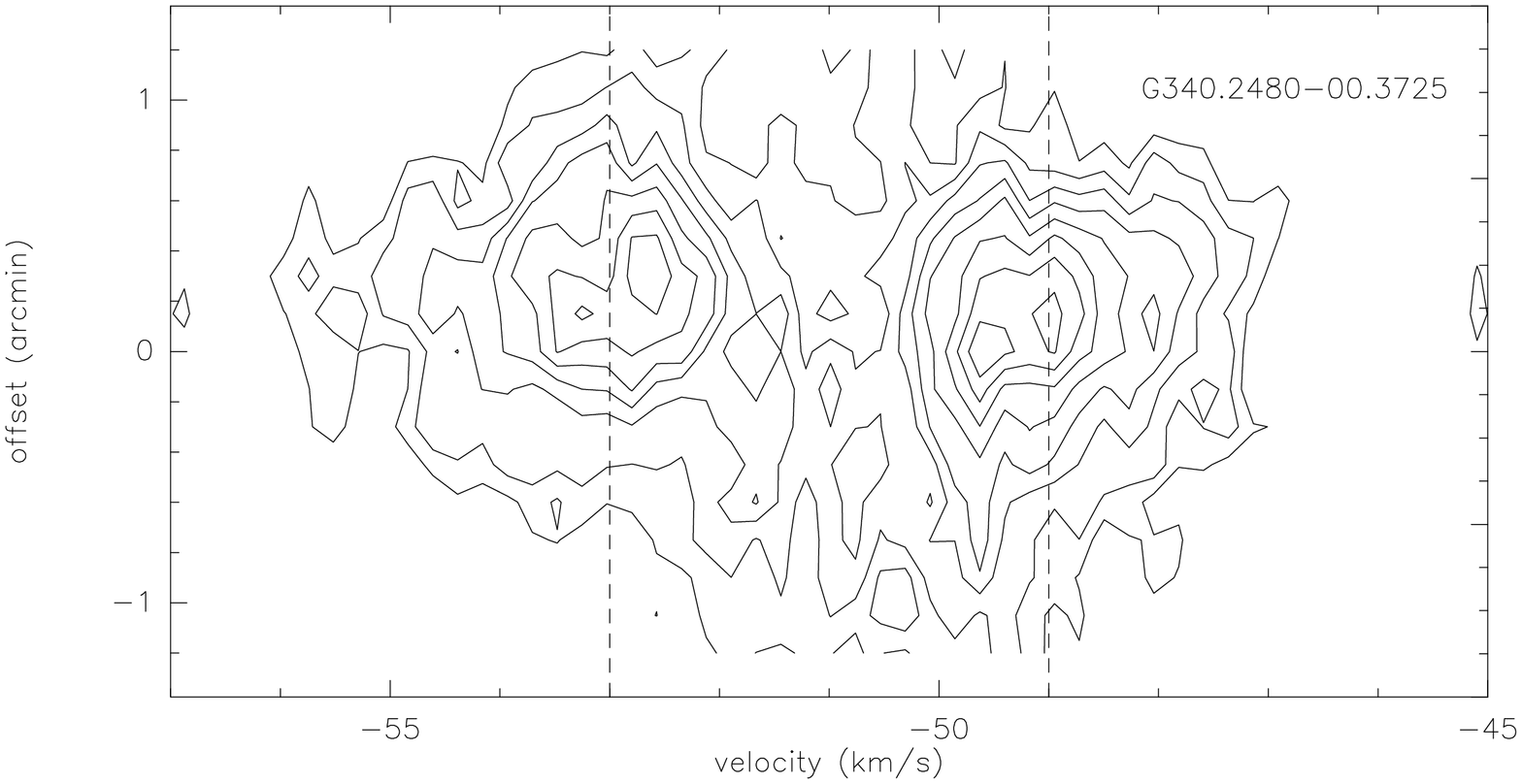}
\includegraphics[width=3in,height=3in,angle=270]{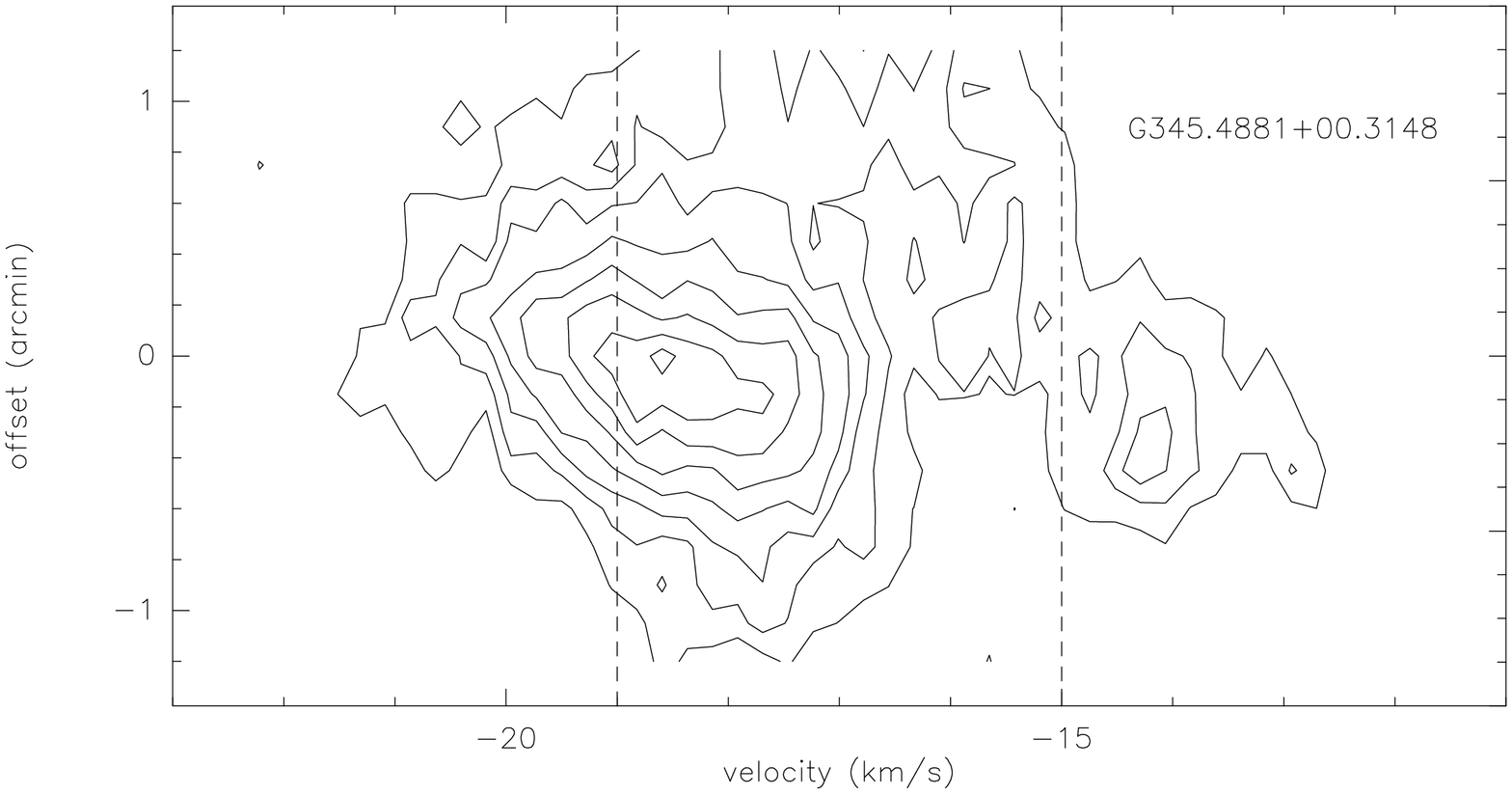}
\caption{ PV diagrams of the HII candidates show HCO$^+$ (1-0) wing
emissions. Contours are 20$\%$, 30$\%$...90$\%$ of the peak
emissions. The dashed lines indicate the velocity ranges for the
blue and red wings as listed in table 2.}
\end{figure*}

\begin{figure*}
\onecolumn \centering
\includegraphics[width=2.5in,height=3.2in,angle=270]{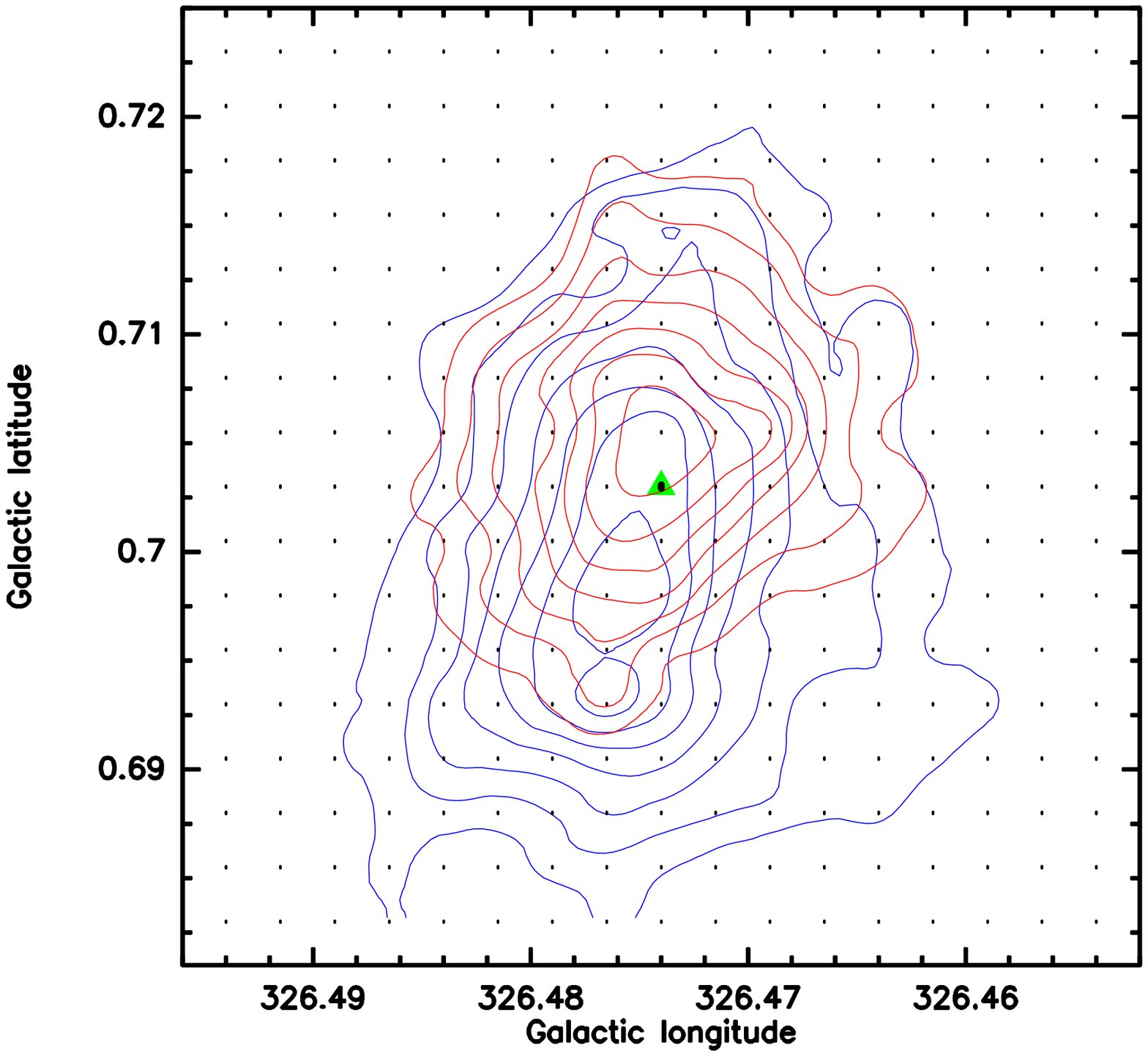}
\includegraphics[width=2.5in,height=3.2in,angle=270]{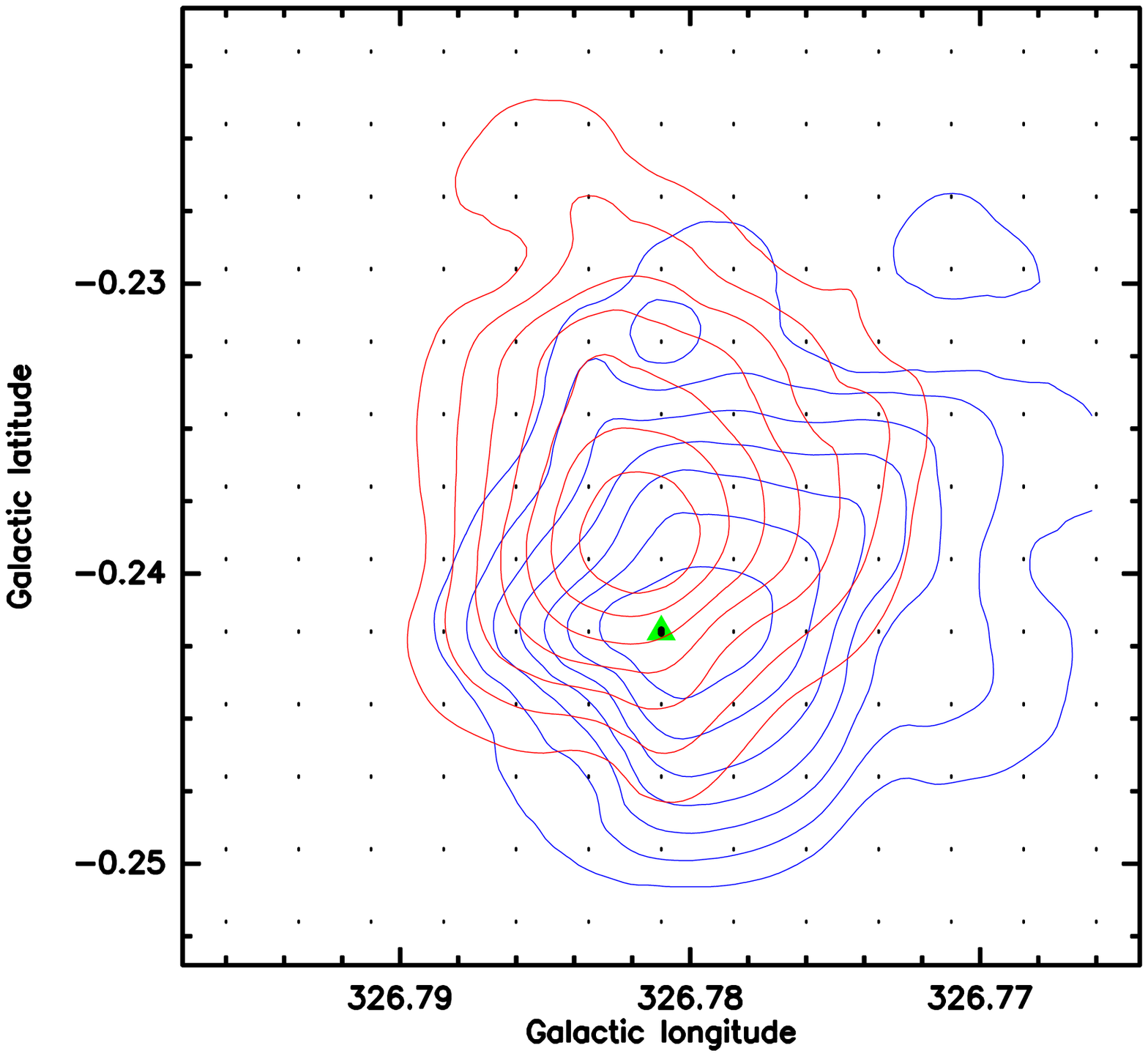}
\includegraphics[width=2.5in,height=3.2in,angle=270]{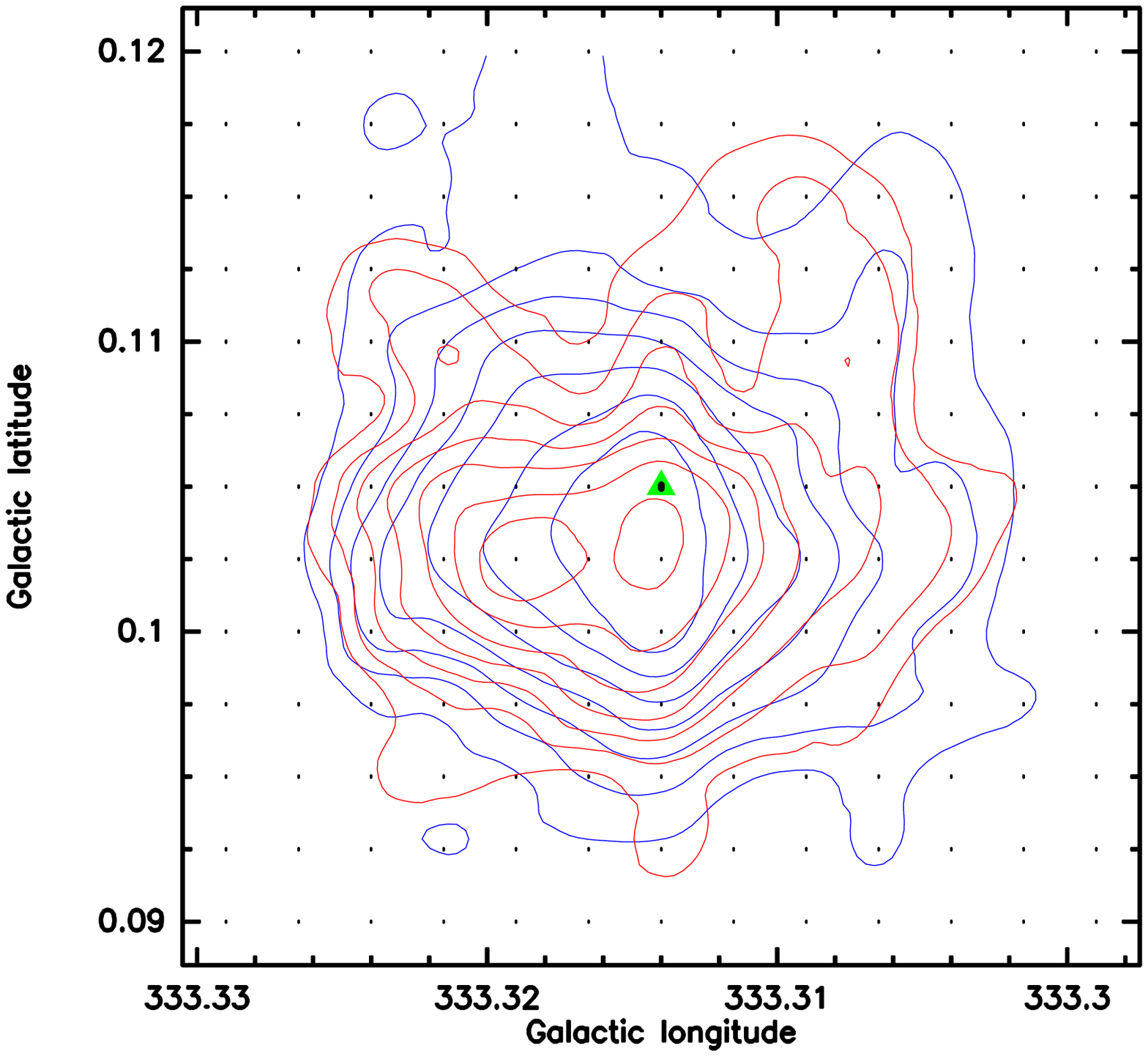}
\includegraphics[width=2.5in,height=3.2in,angle=270]{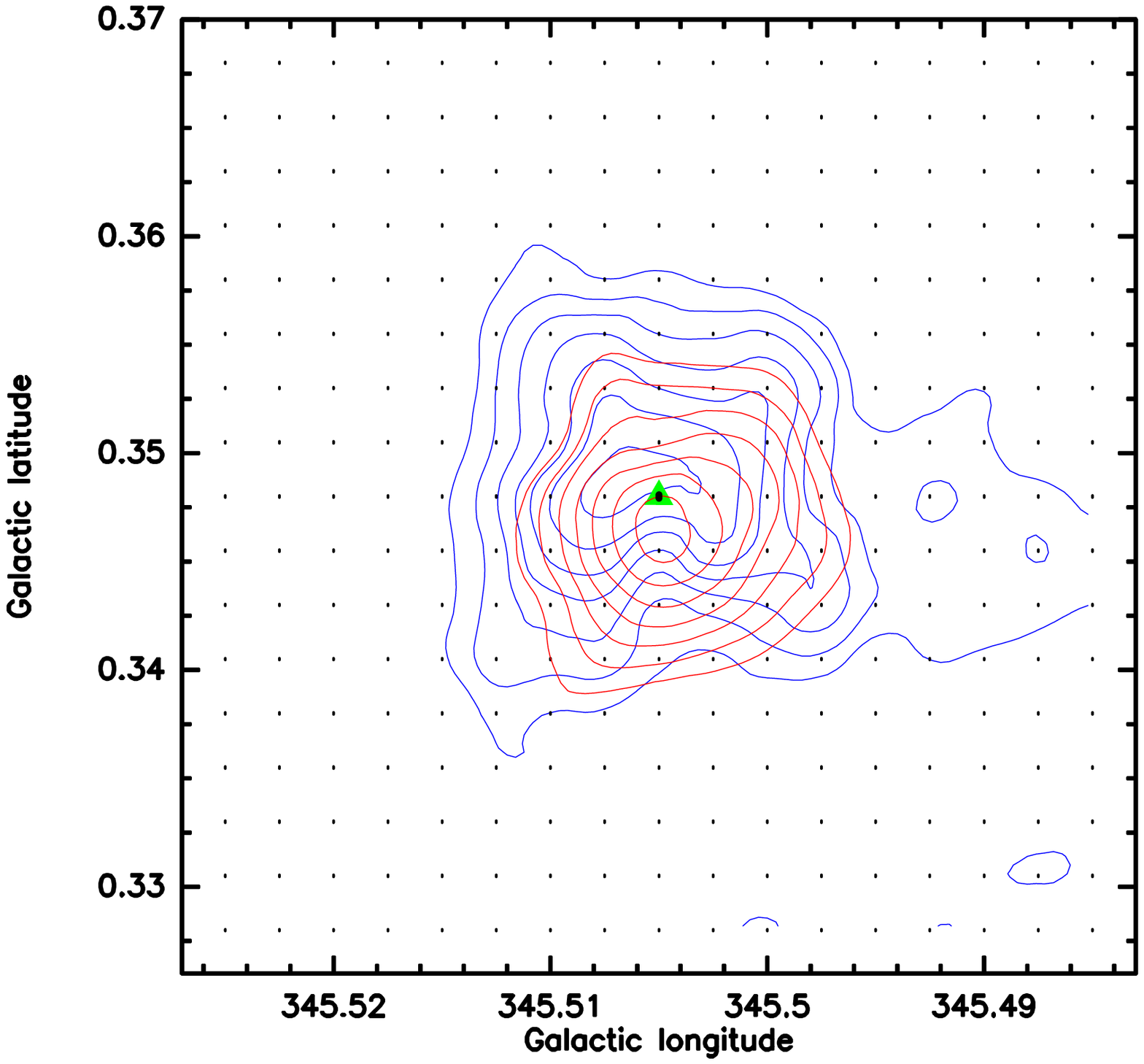}
\caption{The blue and red contours show the HCO$^+$ (1-0) integrated
wing emissions of the MYSO candidates. The green triangles mark the
RMS sources and the dot symbols mark the mapped points. The levels
are 30$\%$, 40$\%$...90$\%$ of the peak emissions respectively. }
\end{figure*}

\begin{figure*}
\onecolumn \centering
\includegraphics[width=2.5in,height=3.2in,angle=270]{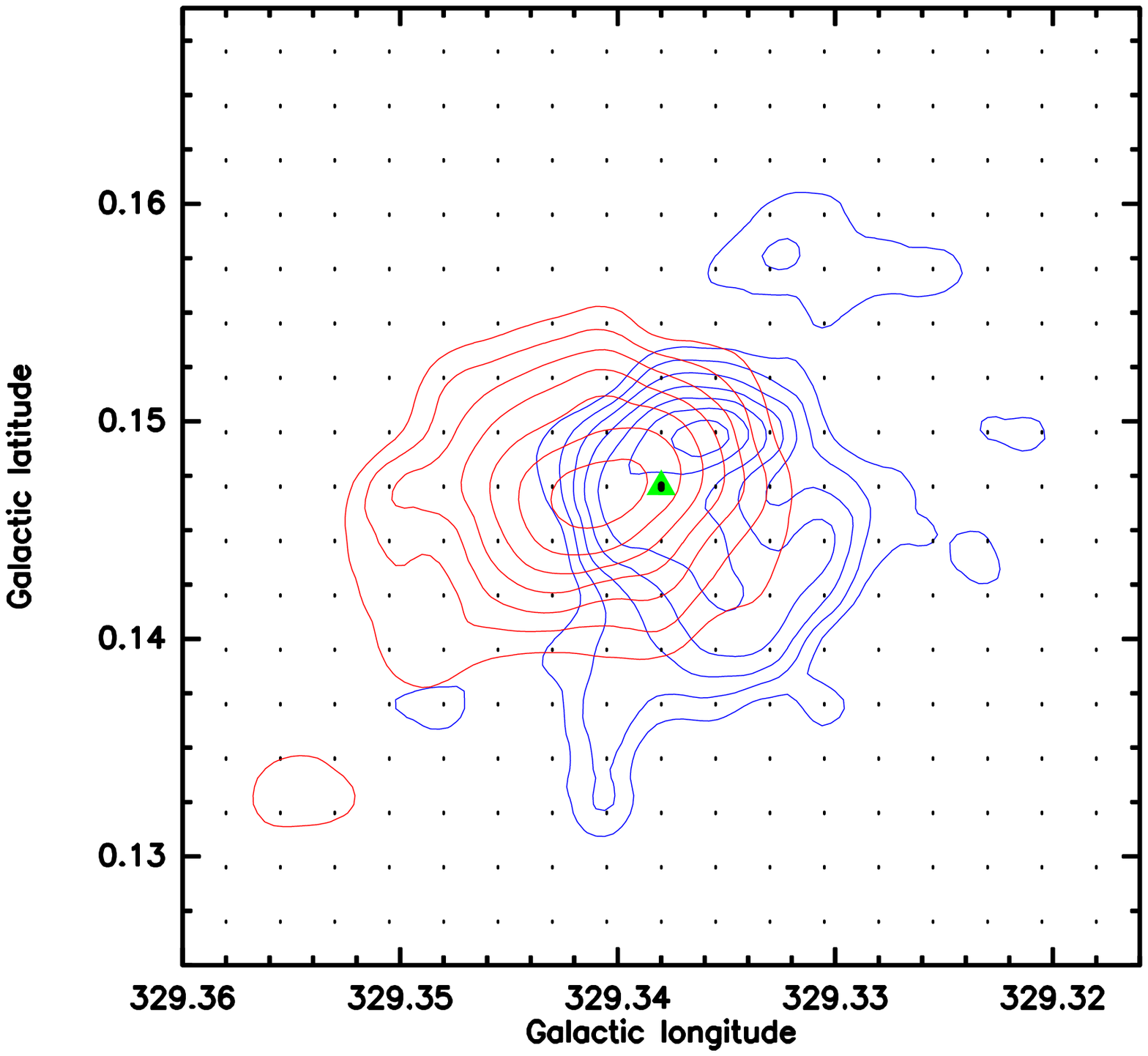}
\includegraphics[width=2.5in,height=3.2in,angle=270]{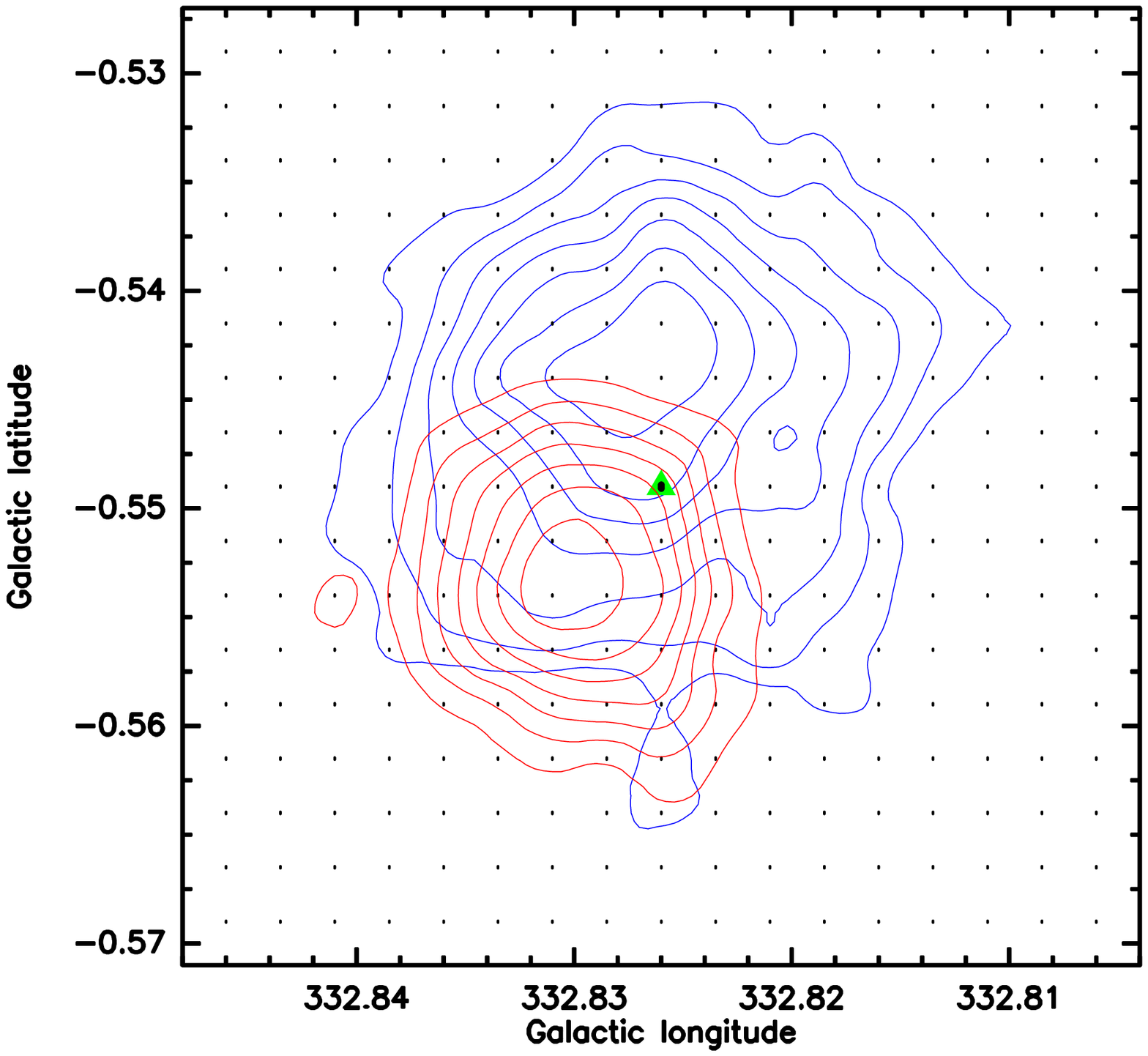}
\includegraphics[width=2.5in,height=3.2in,angle=270]{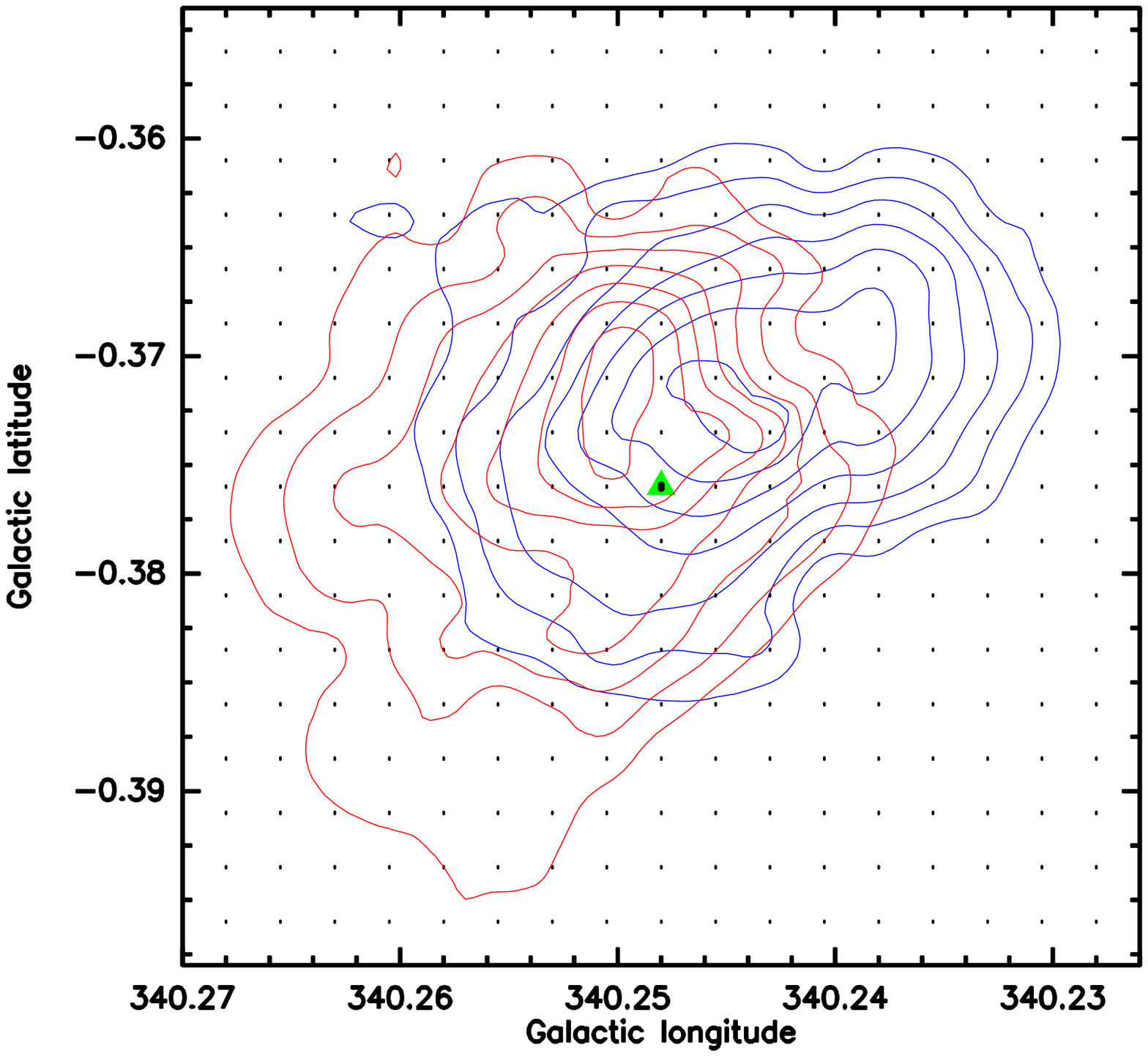}
\includegraphics[width=2.5in,height=3.2in,angle=270]{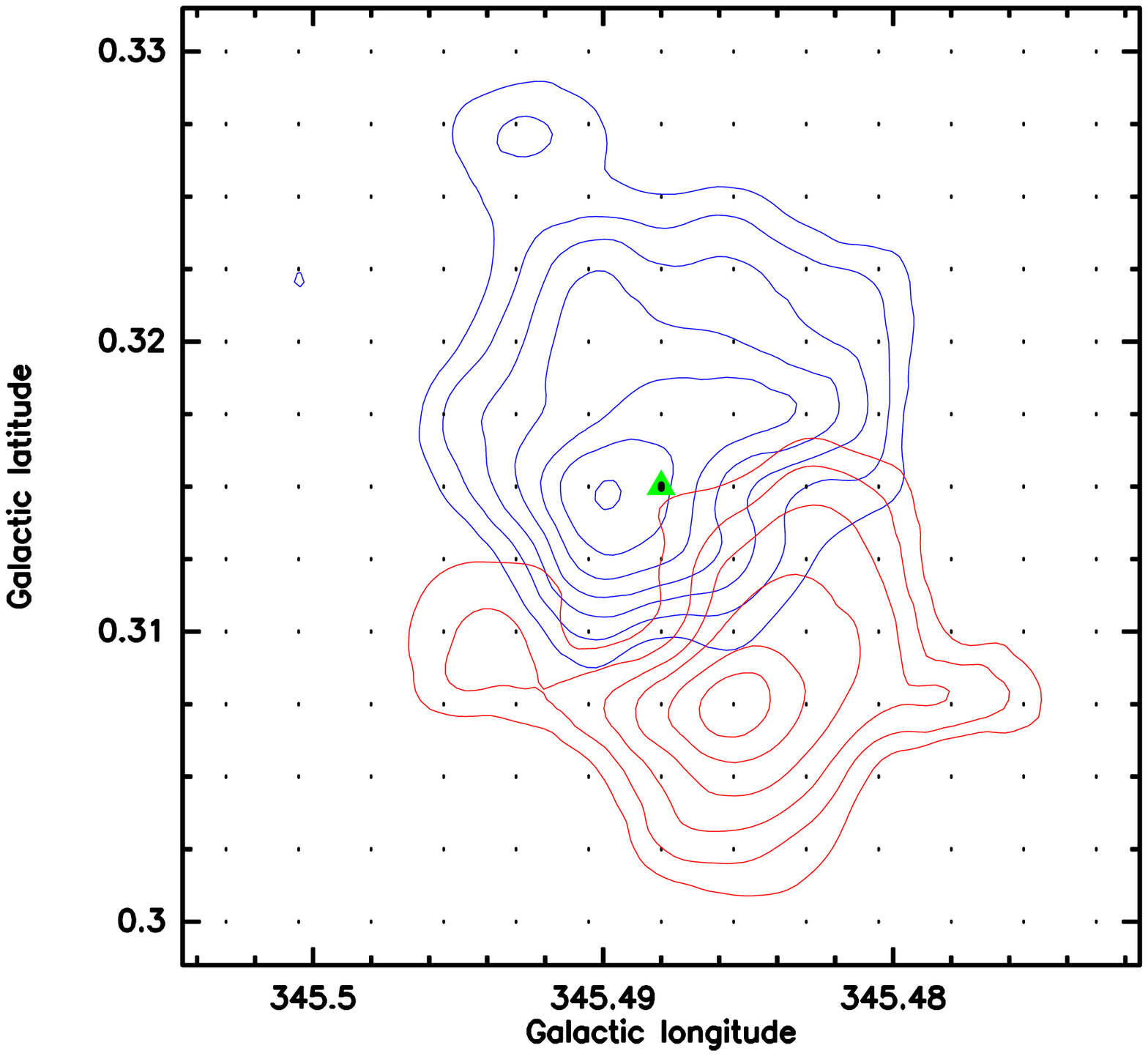}
\caption{ The blue and red contours show the HCO$^+$ (1-0)
integrated wing emissions of the HII candidates. The green triangles
mark the RMS sources and the dot symbols mark the mapped points. The
levels are 30$\%$, 40$\%$...90$\%$ of the peak emissions
respectively. }
\end{figure*}

\begin{figure*}
\onecolumn \centering
\includegraphics[width=3in,height=3in,angle=270]{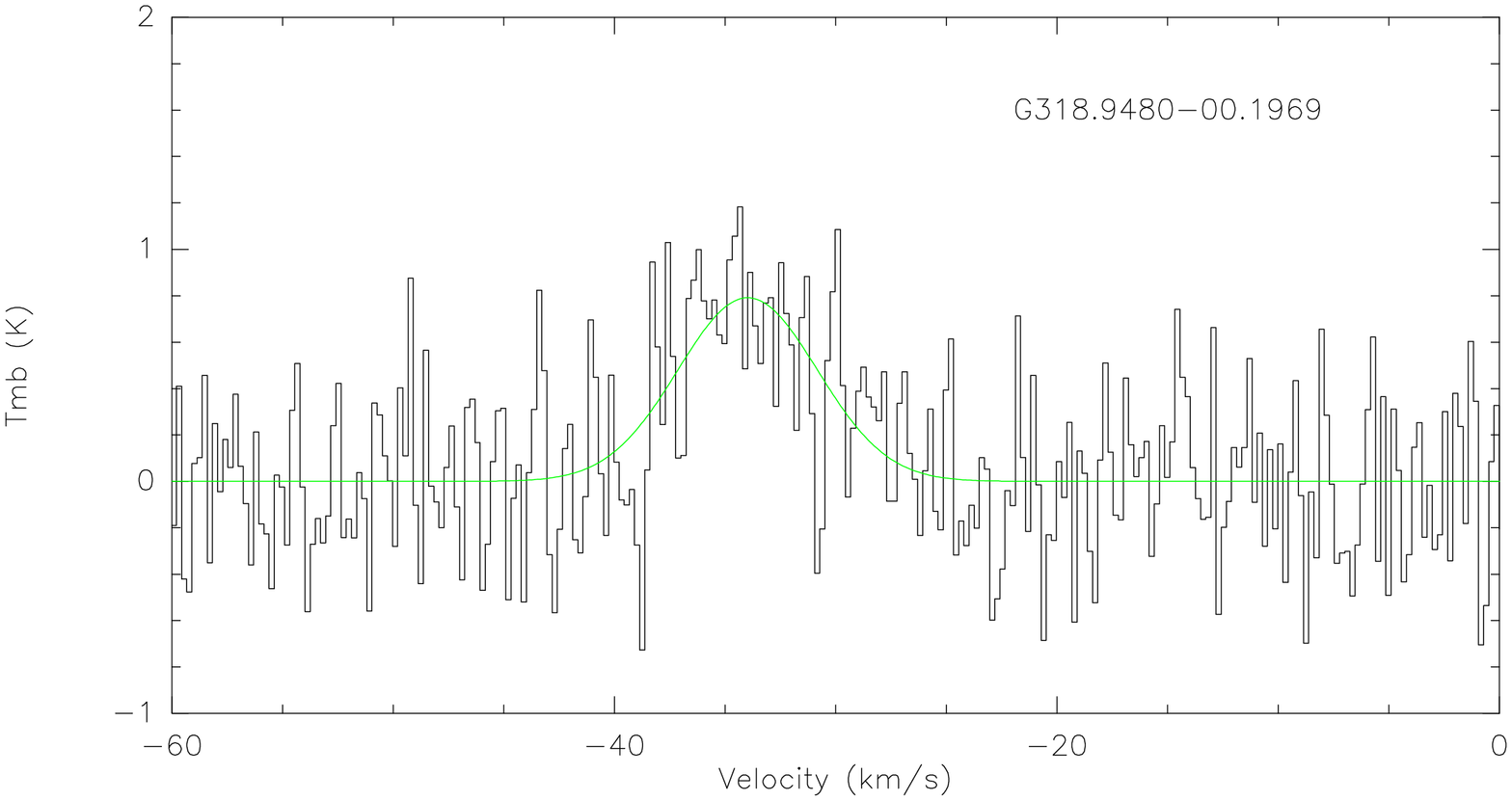}
\includegraphics[width=3in,height=3in,angle=270]{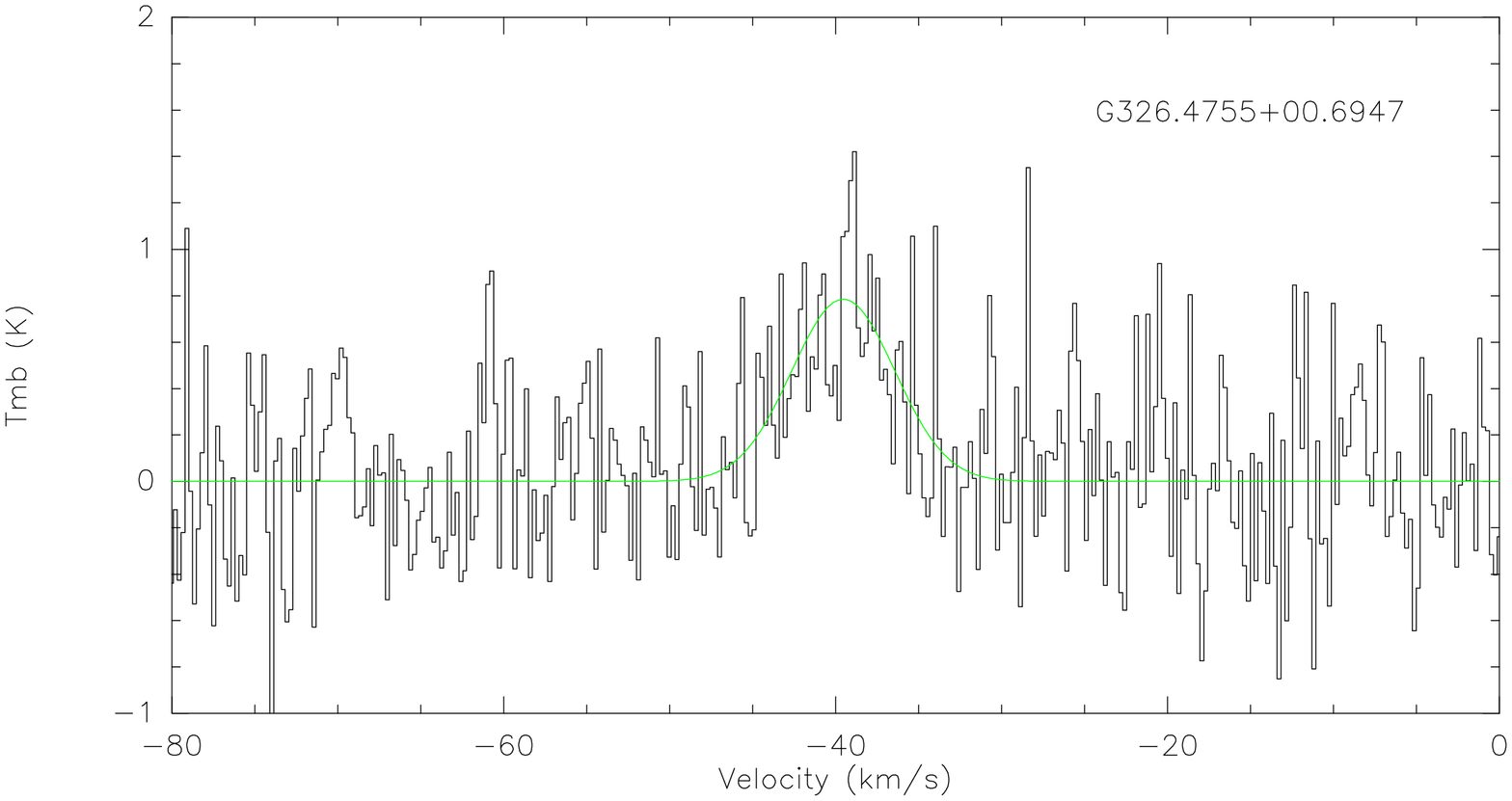}
\includegraphics[width=3in,height=3in,angle=270]{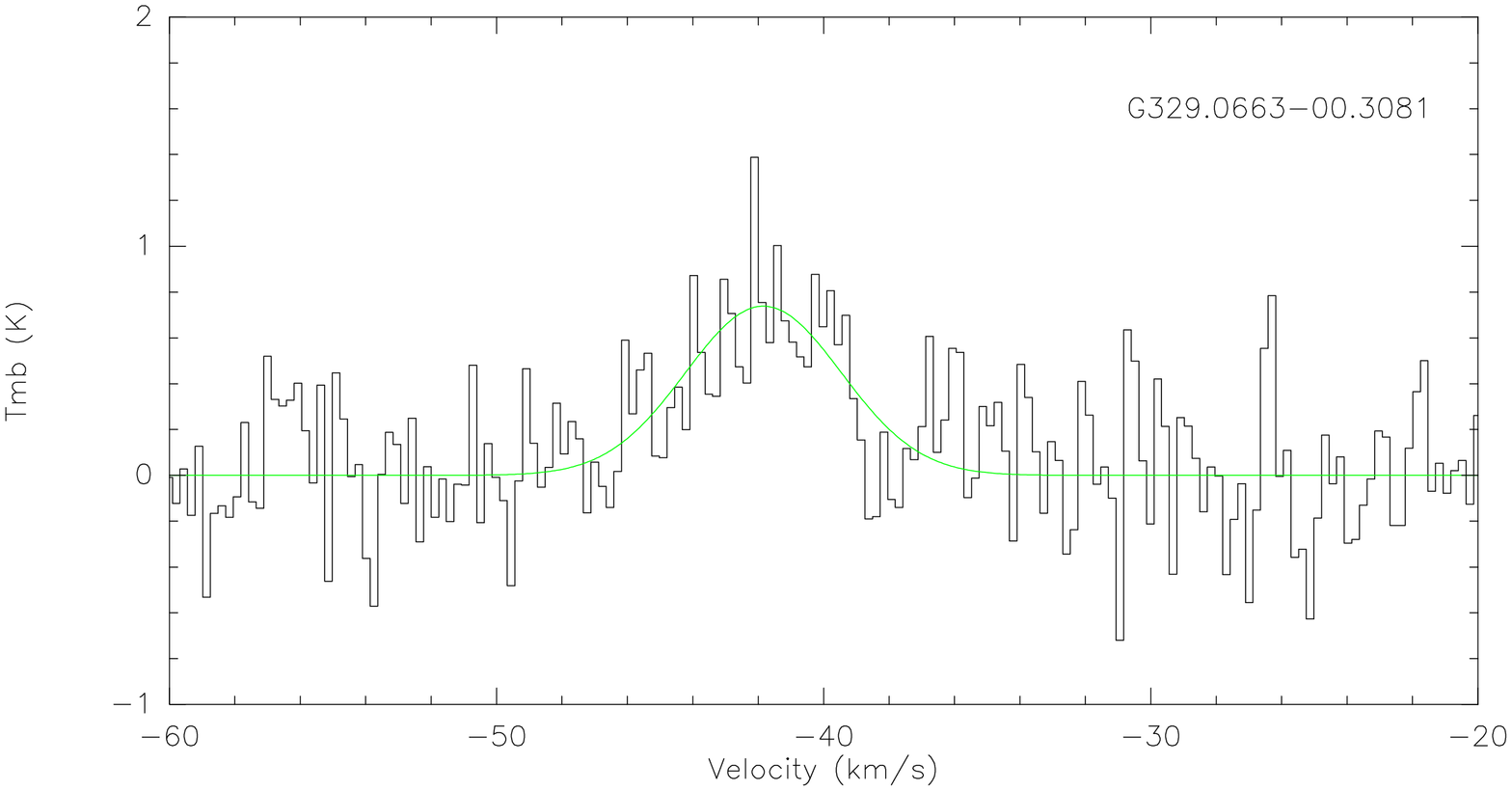}
\includegraphics[width=3in,height=3in,angle=270]{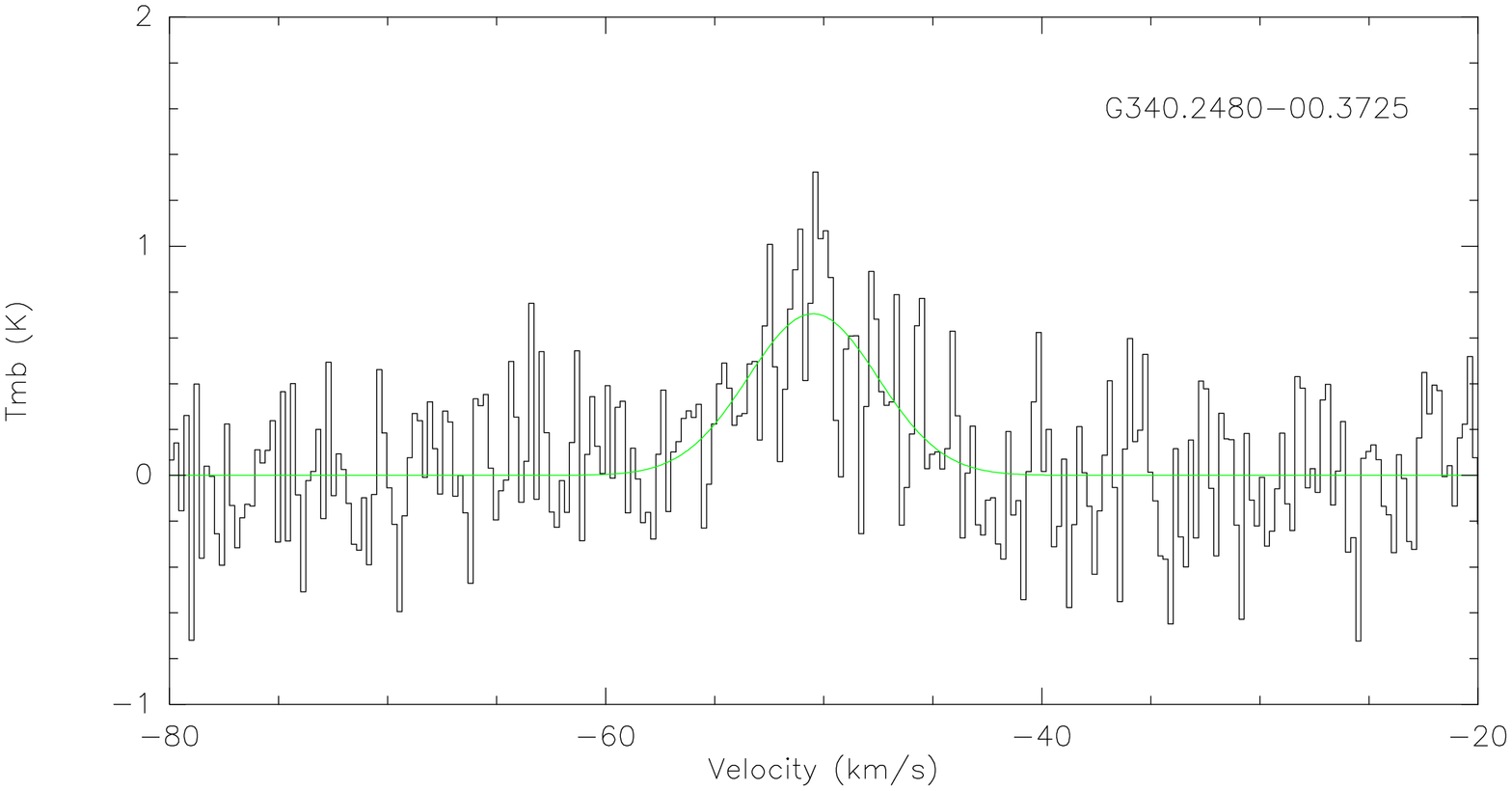}
\includegraphics[width=3in,height=3in,angle=270]{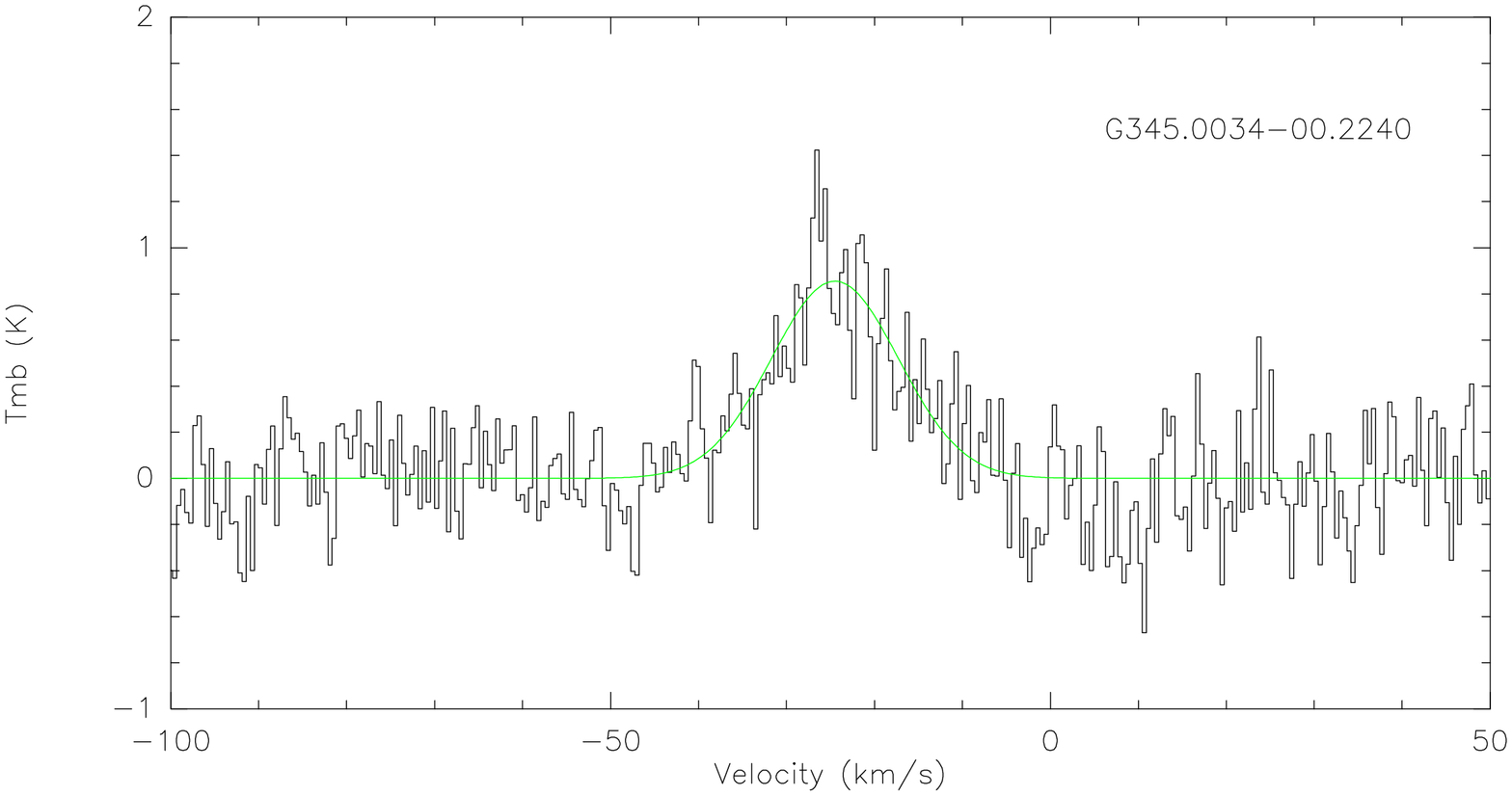}
\caption{ SiO (2-1) spectra for five sources in which SiO was
detected. The green lines are the Gaussian-fitted lines.}
\end{figure*}

\begin{figure*}
\onecolumn \centering
\includegraphics[width=4in,height=4in]{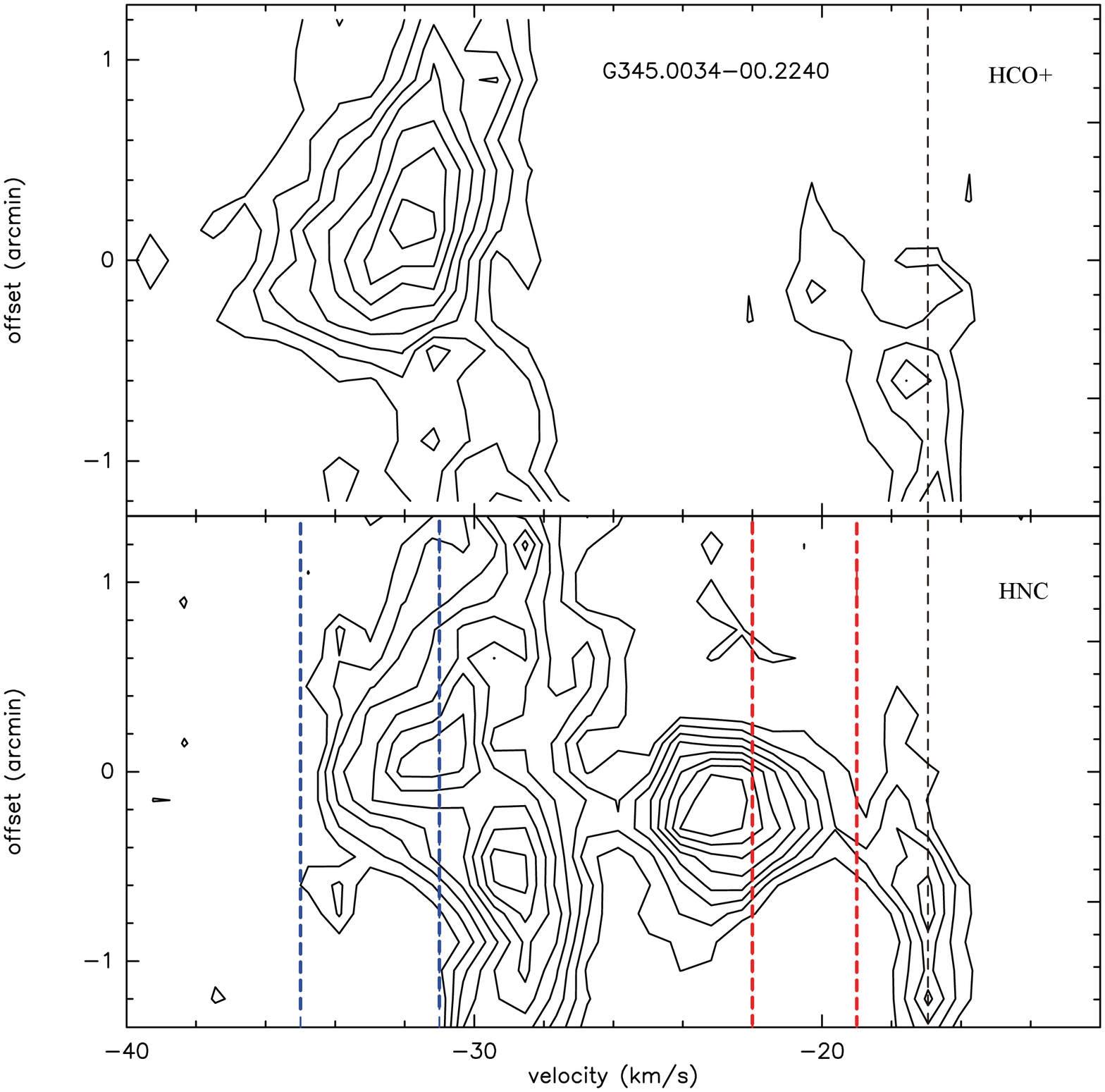}
\caption{ PV diagrams made by HCO$^+$ (top) and HNC (bottom) of
G345.0034-00.2240. Contours are 20$\%$, 30$\%$...90$\%$ of the peak
emissions. The red and blue dashed lines indicate the velocity
ranges for the blue and red wings as listed in table 2. The gas at
-17 km/s is possibly unrelated to this source.}
\end{figure*}

\begin{figure*}
\onecolumn \centering
\includegraphics[width=3in,height=3in,angle=270]{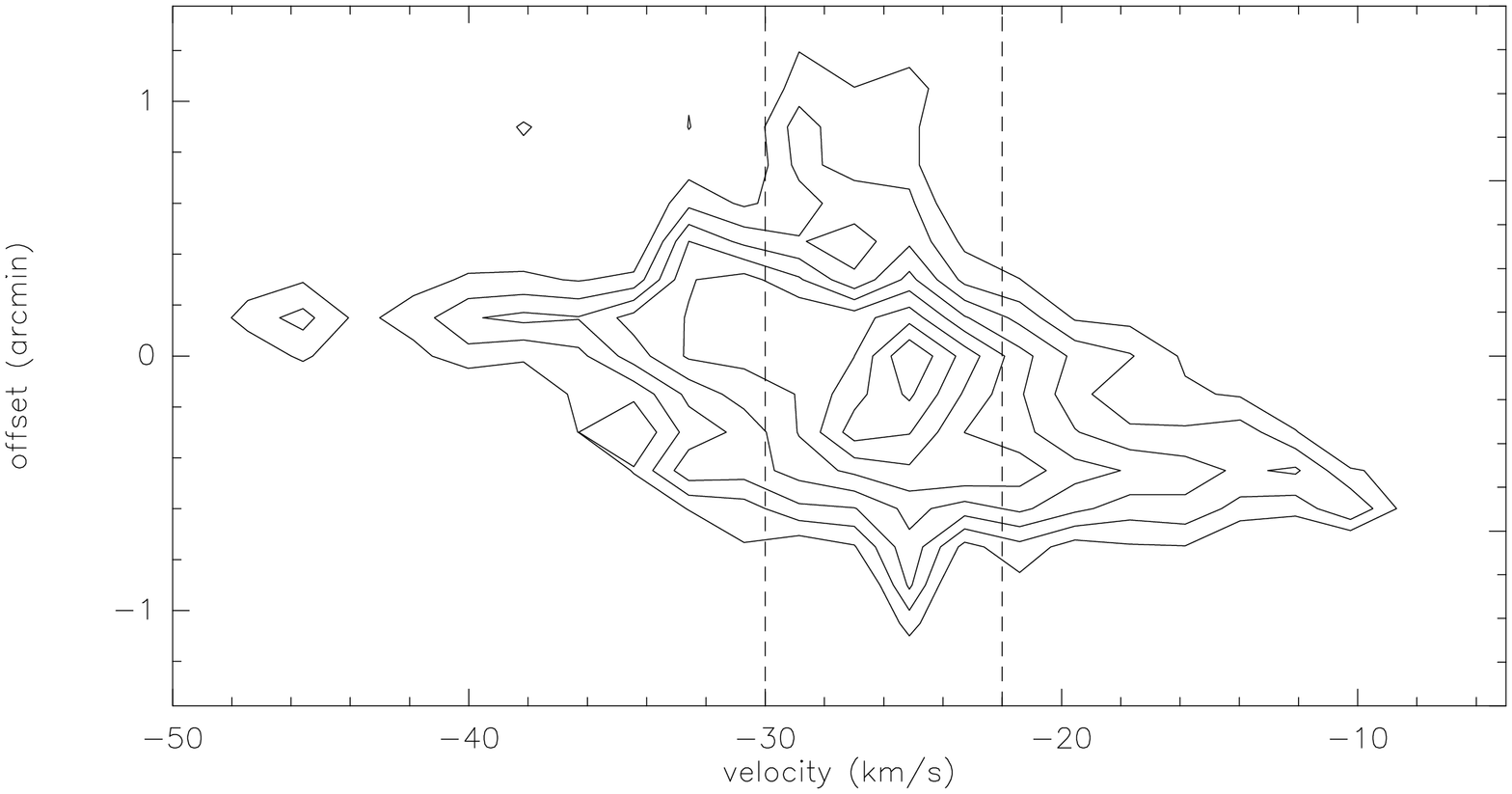}
\includegraphics[width=3in,height=3.5in,angle=270]{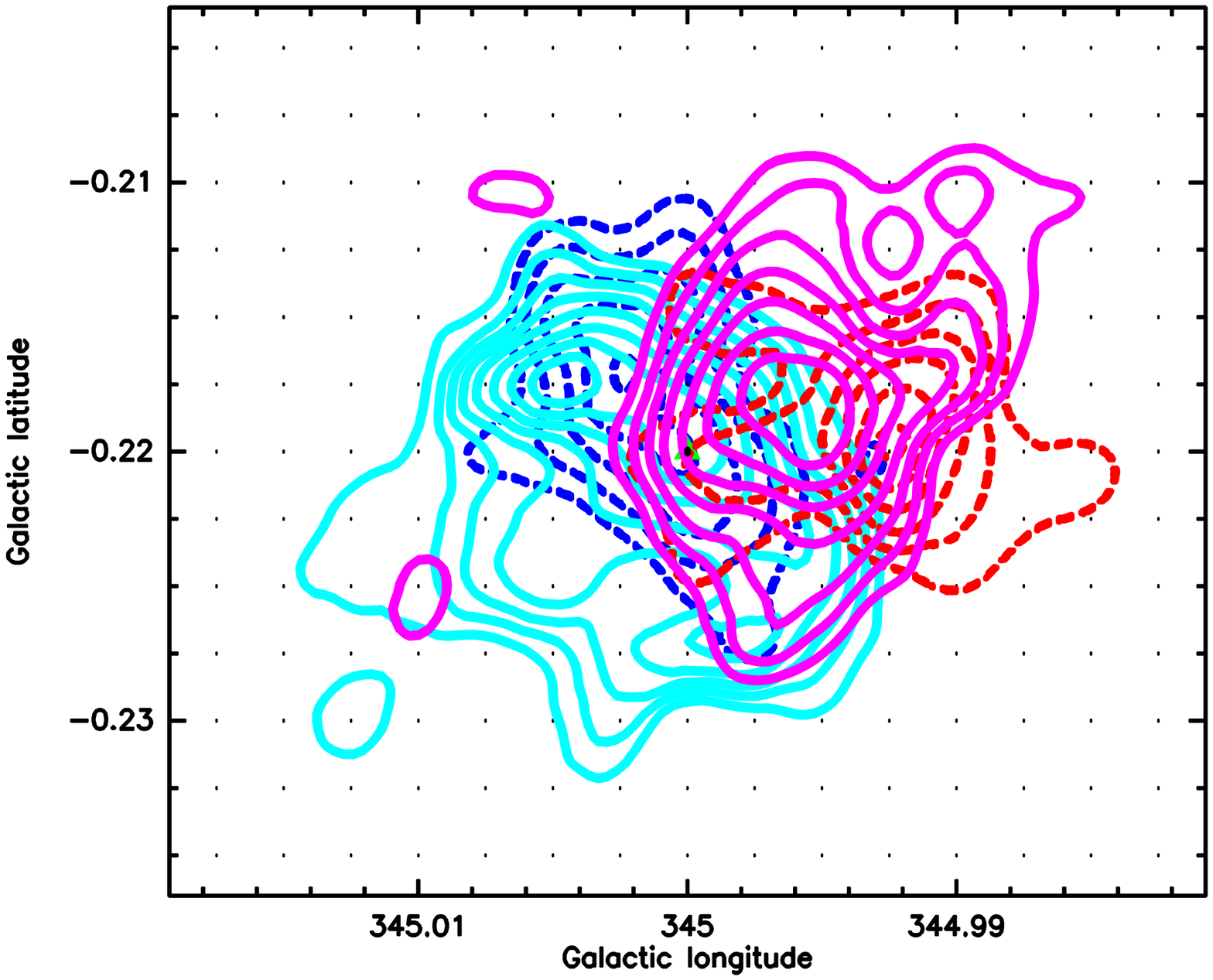}
\caption{ Left: SiO PV diagram of G345.0034-00.2240 cut along
east-west direction. Contour levels are 30$\%$, 40$\%$...90$\%$ of
the peak emission. Right: Outflow traced by SiO (dashed red and blue
contours) superimposed with HNC$^+$ wing emissions (cyan and pink
contours). }
\end{figure*}

\begin{figure*}
\onecolumn \centering
\includegraphics[width=1.8in,height=3in,angle=270]{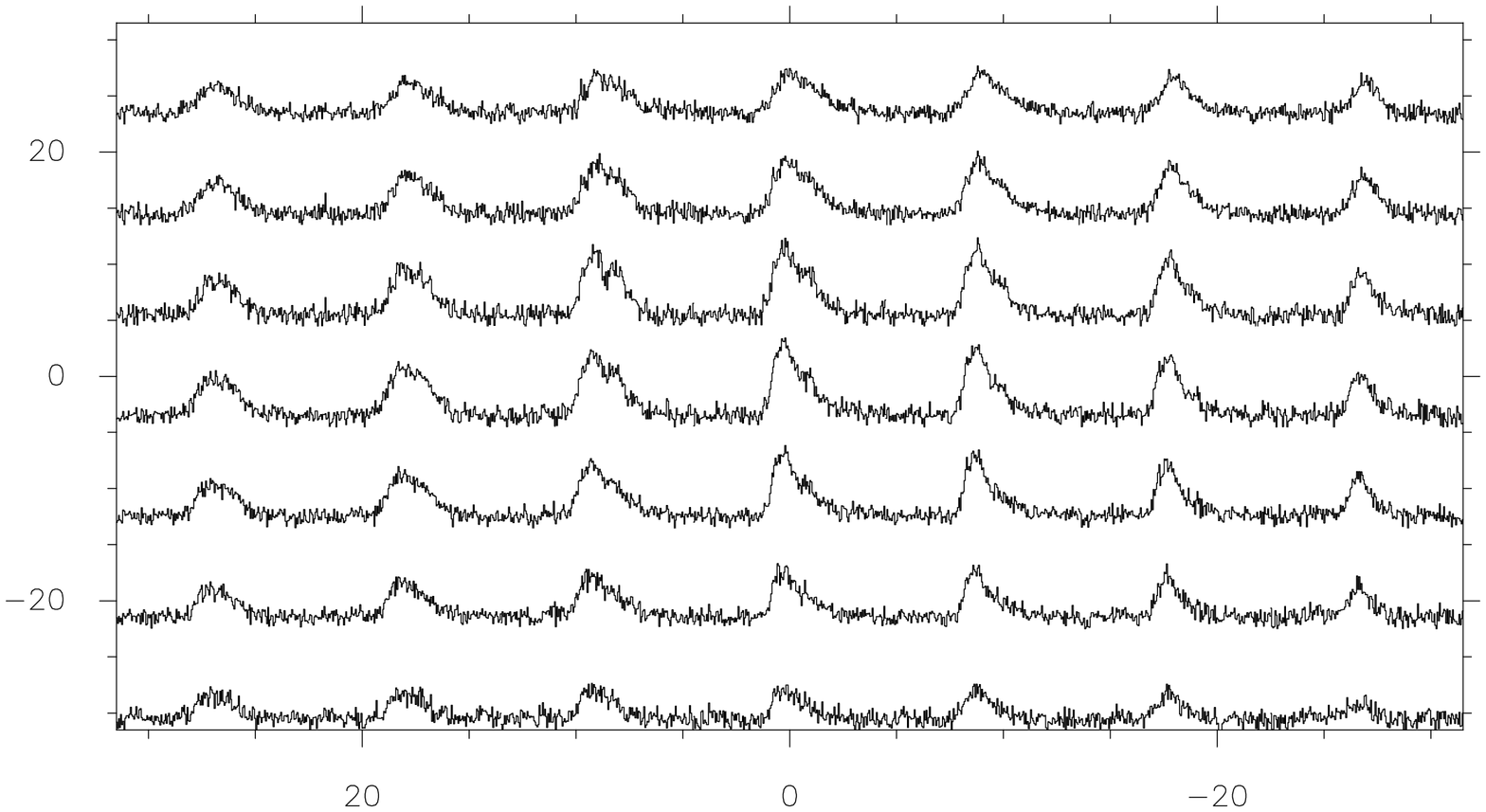}
\includegraphics[width=1.8in,height=3in,angle=270]{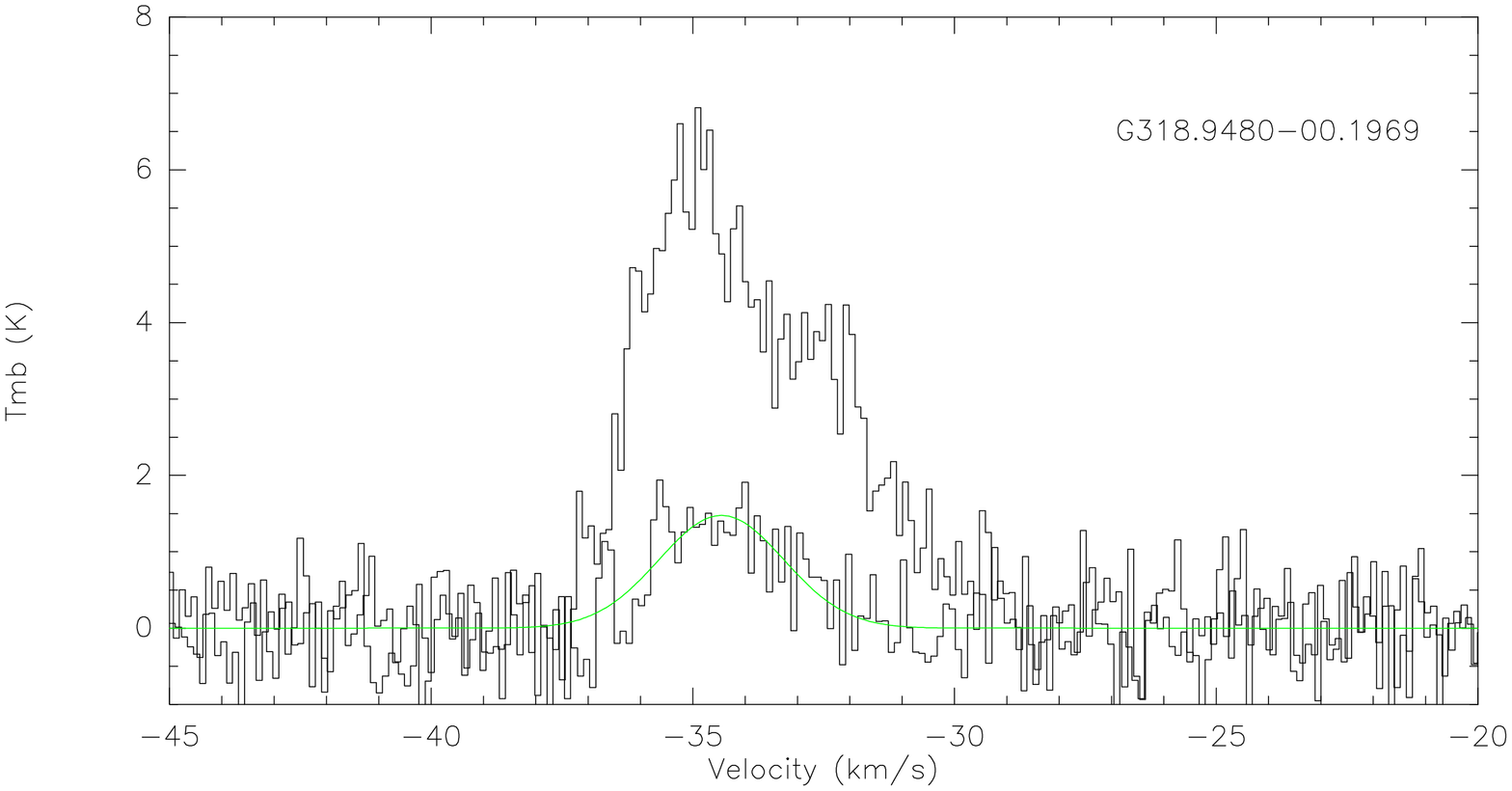}
\includegraphics[width=1.8in,height=3in,angle=270]{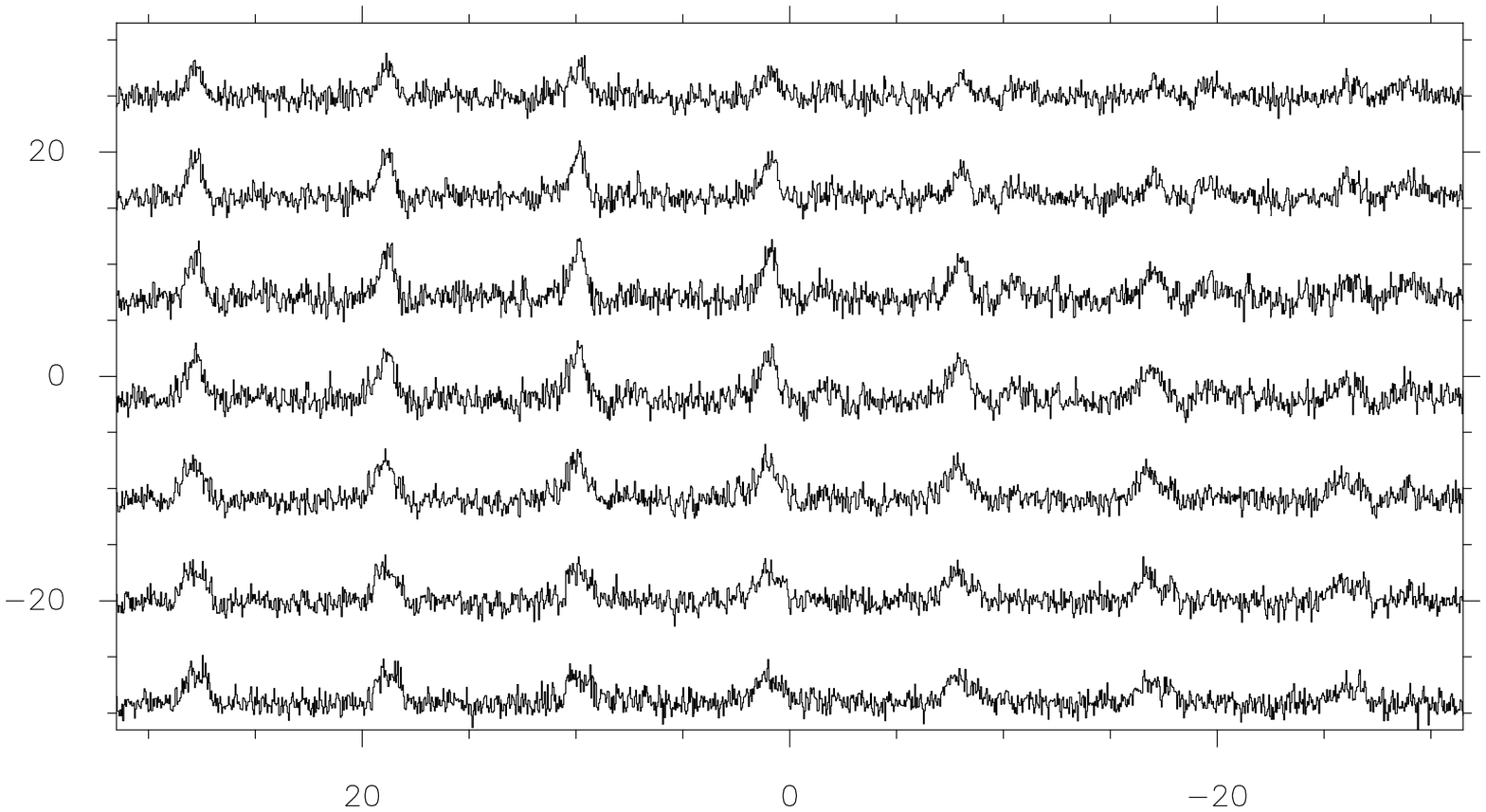}
\includegraphics[width=1.8in,height=3in,angle=270]{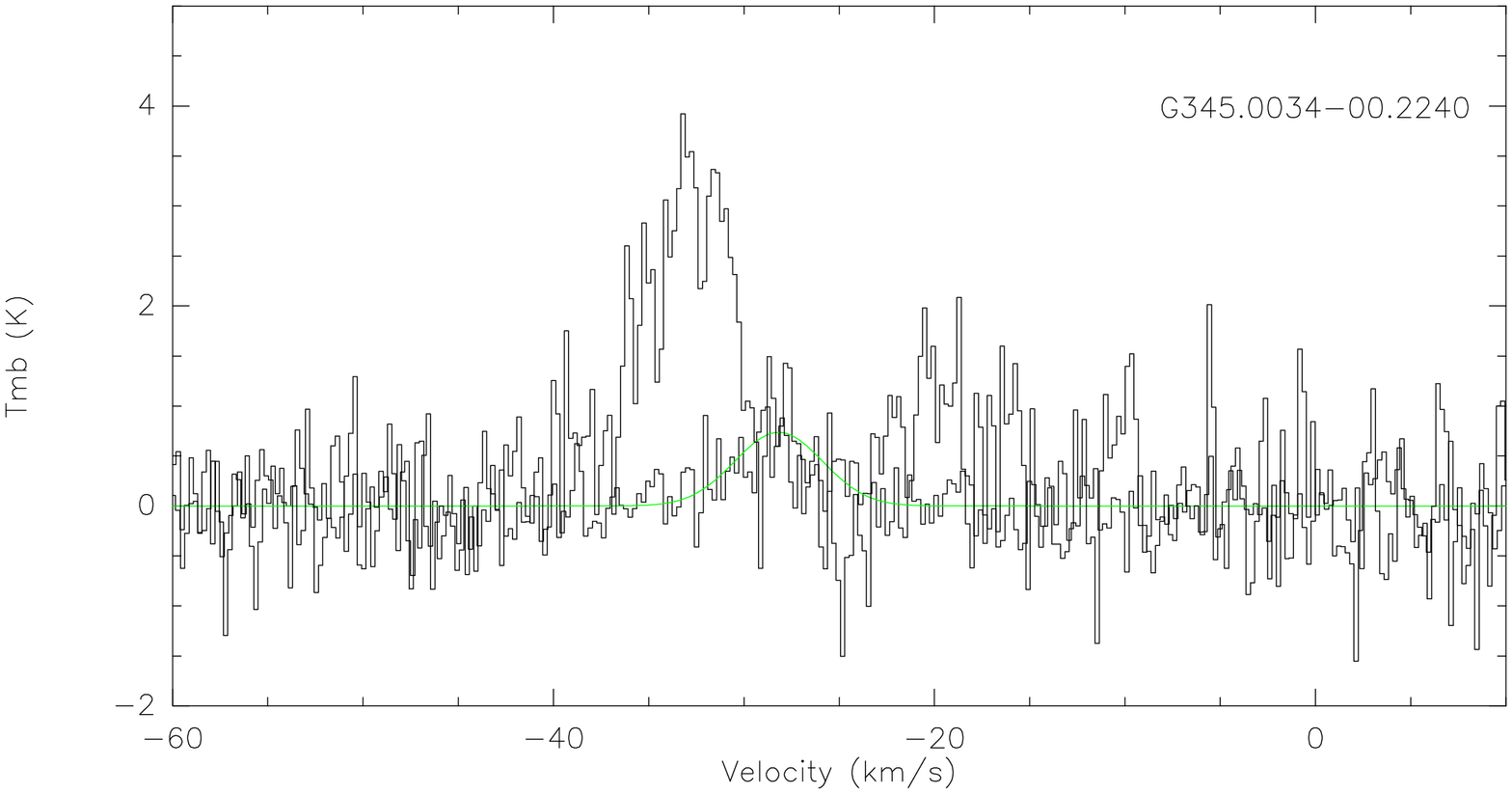}
\includegraphics[width=1.8in,height=3in,angle=270]{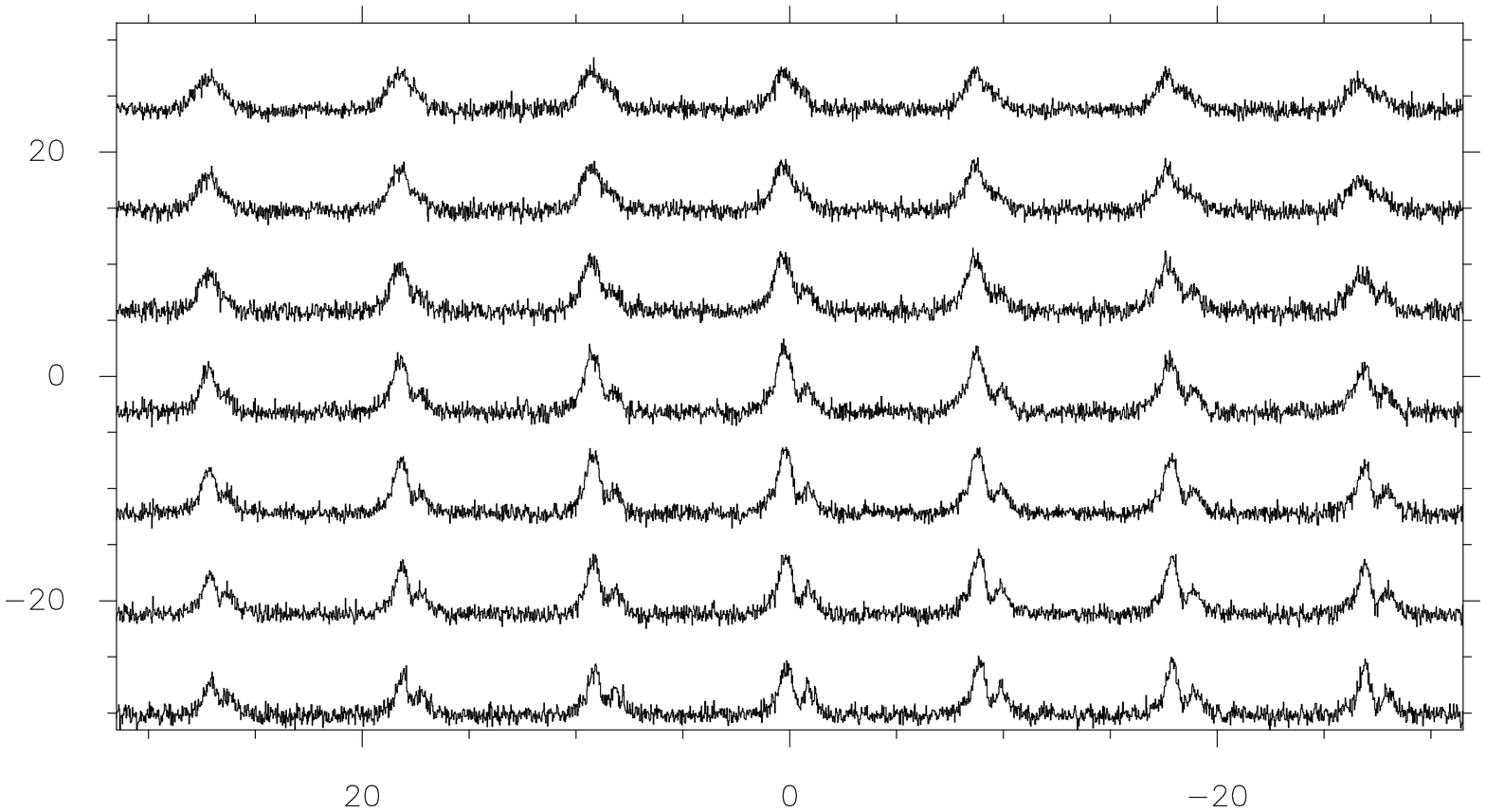}
\includegraphics[width=1.8in,height=3in,angle=270]{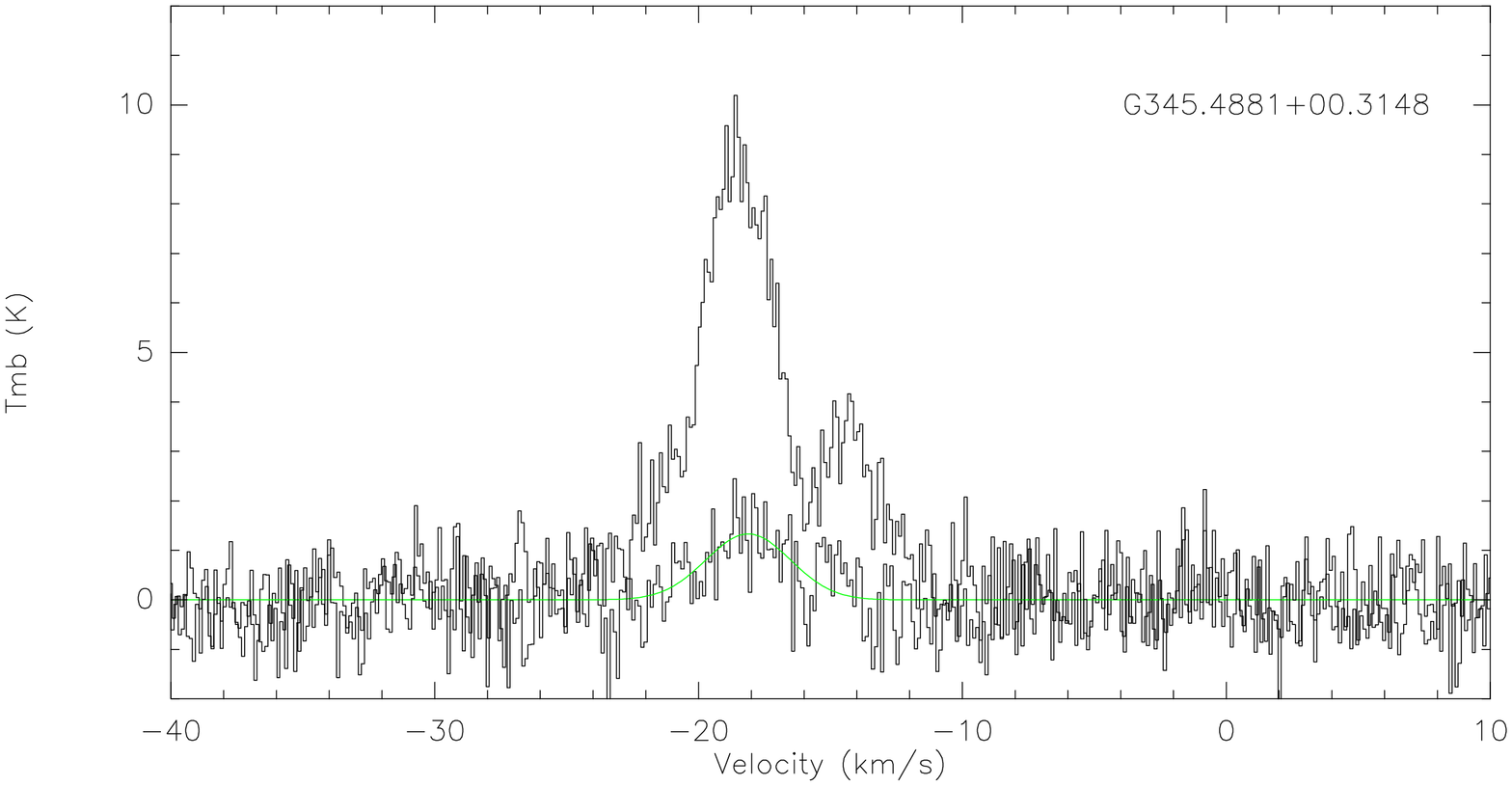}
\includegraphics[width=1.8in,height=3in,angle=270]{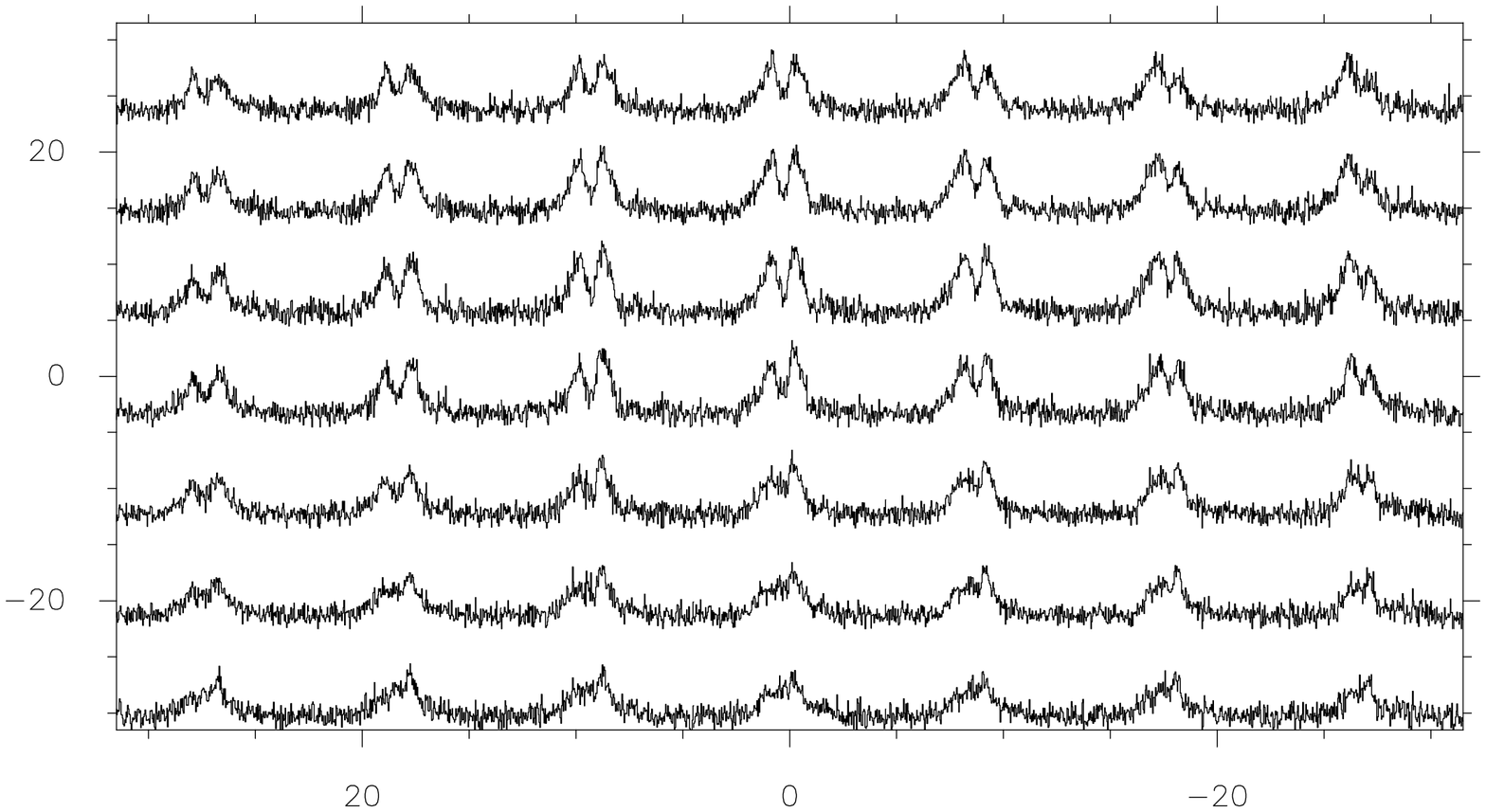}
\includegraphics[width=1.8in,height=3in,angle=270]{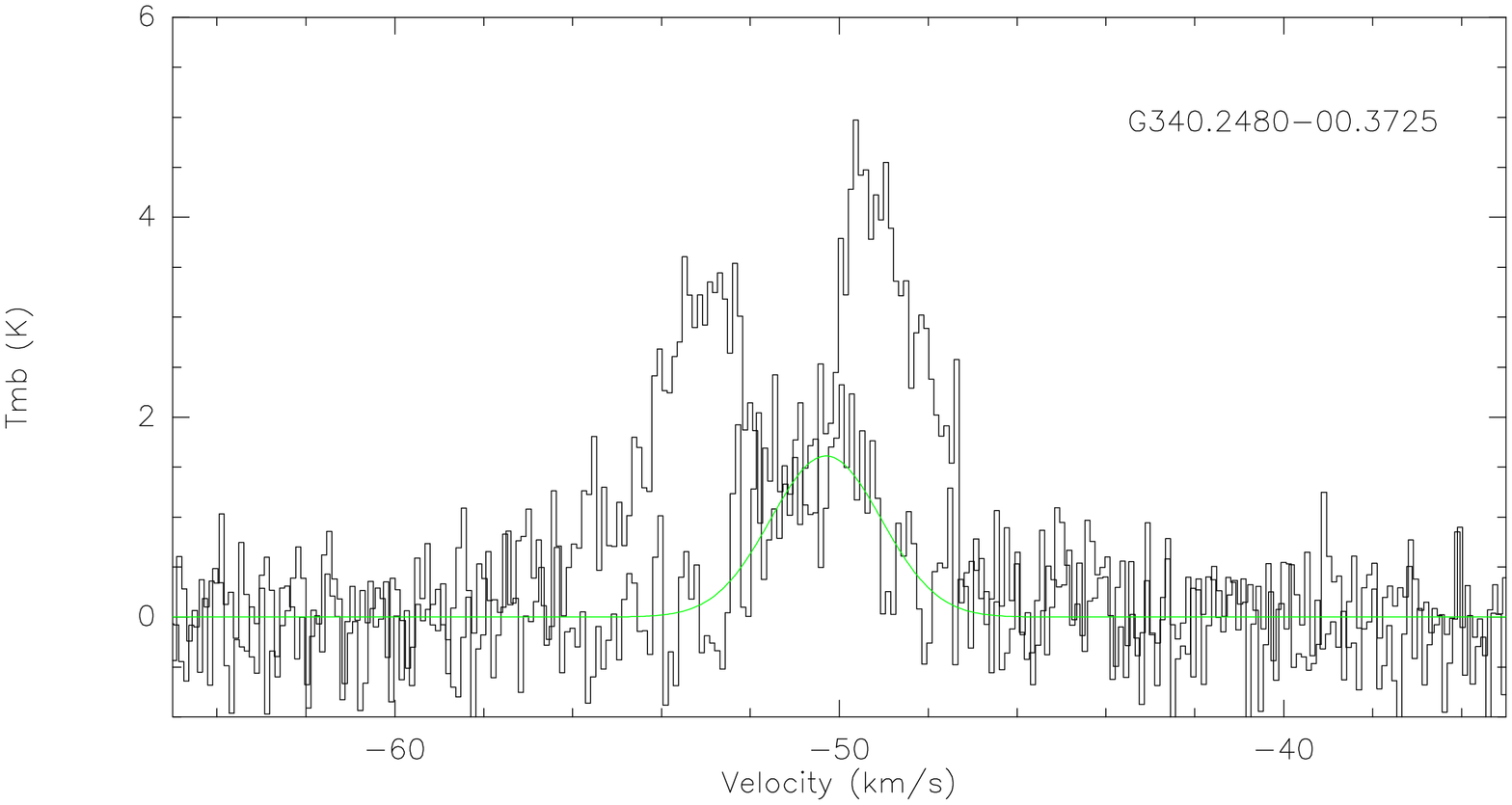}
\includegraphics[width=1.8in,height=3in,angle=270]{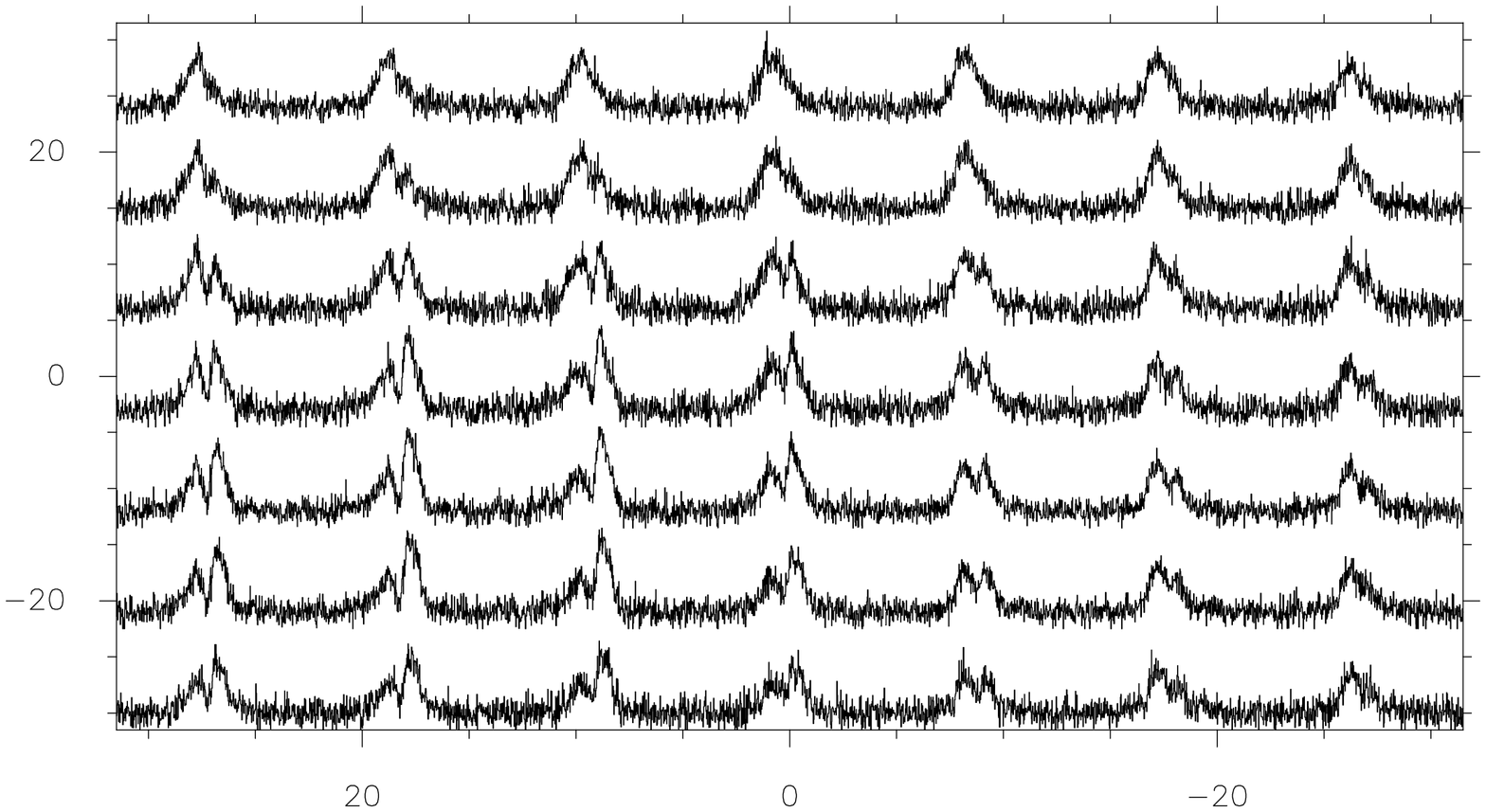}
\includegraphics[width=1.8in,height=3in,angle=270]{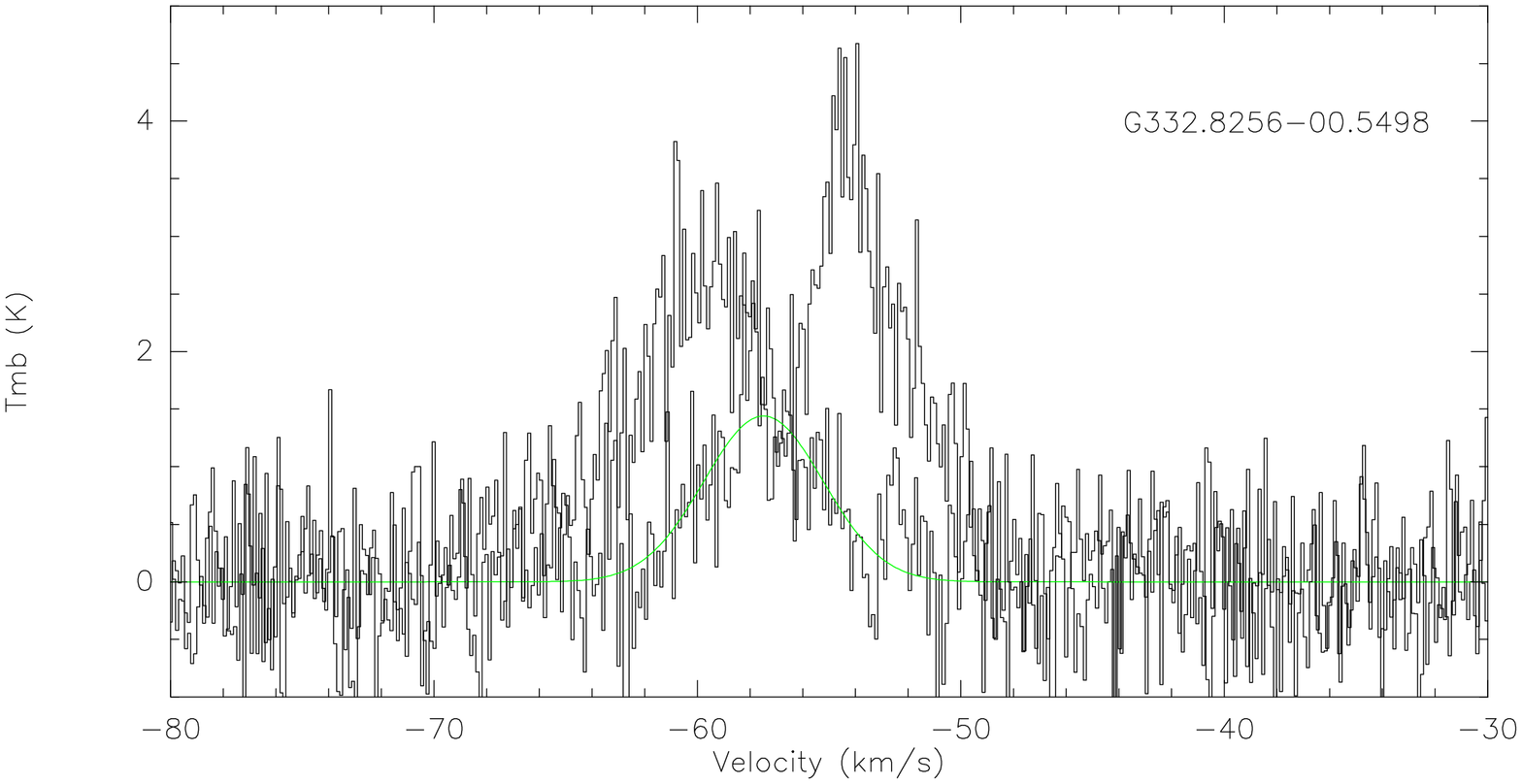}
\caption{From the top and bottom panels present G318.9480-00.1969,
G345.0034-00.2240, G345.4881+00.3148, G340.2480-00.3725 and
G332.8256-00.5498 HCO$^+$ mapping observations (left) and its
central spectra of HCO$^+$ and H$^{13}$CO$^+$ (right). The green
lines are the H$^{13}$CO$^+$ Gaussian-fitted lines. }
\end{figure*}

\begin{figure*}
\onecolumn \centering
\includegraphics[width=4in,height=3in]{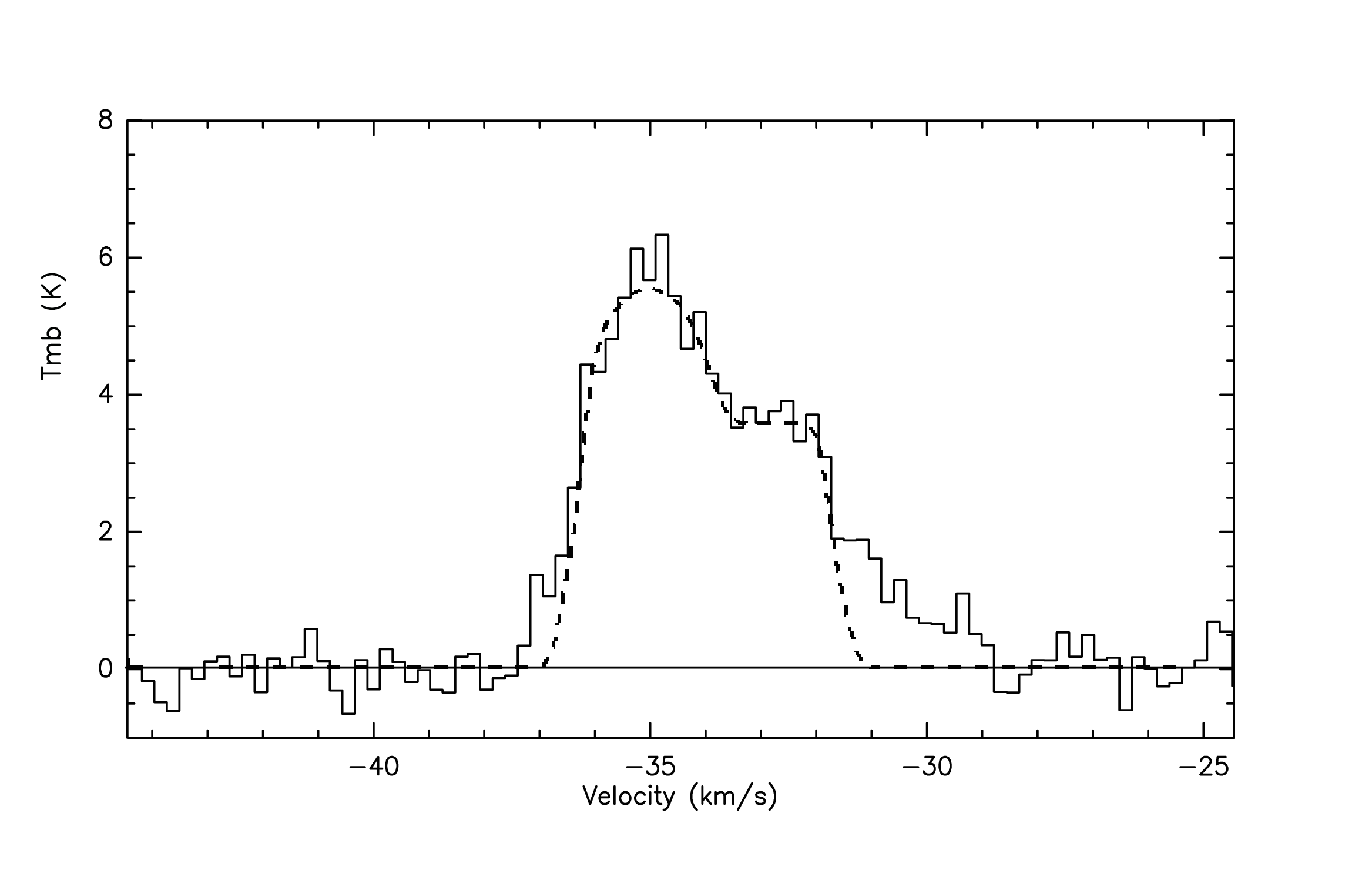}
\caption{A comparison of the observed (solid) and model (dashed)
HCO$^+$ spectra of G318.9480-00.1969. }
\end{figure*}

\begin{figure*}
\onecolumn \centering
\includegraphics[width=3in,height=3in]{1.sed.eps}
\includegraphics[width=3in,height=3in]{2.sed.eps}
\includegraphics[width=3in,height=3in]{3.sed.eps}
\includegraphics[width=3in,height=3in]{4.sed.eps}
\includegraphics[width=3in,height=3in]{5.sed.eps}
\includegraphics[width=3in,height=3in]{6.sed.eps}
\caption{ The SED fitting models. The dashed line represents the
stellar photosphere model. The black line represents the
best-fitting SED, and the gray lines represent all the other
acceptable YSO fits.}
\end{figure*}

\label{lastpage}
\end{document}